\documentclass[11pt,a4paper]{article}

\usepackage[T1]{fontenc}
\usepackage[utf8]{inputenc}
\usepackage{lmodern}
\usepackage{microtype}

\usepackage[a4paper,margin=1in]{geometry}
\usepackage{setspace}
\usepackage{amsmath}
\usepackage{amssymb}
\usepackage{amsthm}
\usepackage{mathtools}
\usepackage{mathrsfs}

\usepackage{graphicx}
\usepackage{float}
\usepackage{booktabs}
\usepackage{tabularx}
\usepackage{array}

\newcolumntype{C}{>{\centering\arraybackslash}X}

\usepackage{enumitem}

\usepackage[numbers,round]{natbib}

\usepackage[colorlinks=true,
            linkcolor=blue,
            citecolor=blue,
            urlcolor=blue,
            bookmarksnumbered=true]{hyperref}

\theoremstyle{plain}

\newtheorem{Theorem}{Theorem}[section]
\newtheorem{Lemma}[Theorem]{Lemma}
\newtheorem{Proposition}[Theorem]{Proposition}
\newtheorem{Corollary}[Theorem]{Corollary}

\theoremstyle{definition}

\newtheorem{Definition}[Theorem]{Definition}
\newtheorem{Example}[Theorem]{Example}

\theoremstyle{remark}

\newtheorem{Remark}[Theorem]{Remark}

\title{\textbf{Semisimple--Unipotent Decomposition and the
Exact Order of a Linear Cellular Automaton over
$\mathbb{F}_2$}}

\author{
Ana Boji\'c\\
\small Independent Researcher\\
\small \texttt{bojicana12@gmail.com}
}

\date{}

\begin{document}

\maketitle

\begin{abstract}
We study the linear cellular automaton on the cyclic graph
$C_n$ over $\mathbb{F}_2$ defined by the local rule
$v_i(t+1)=v_{i-1}(t)\oplus v_i(t)\oplus v_{i+1}(t)$
(Wolfram's Rule 150 under periodic boundary conditions).
Identifying the configuration space $V_n=\mathbb{F}_2^n$
with $R_n = \mathbb{F}_2[x] / \langle x^n - 1 \rangle$,
the evolution operator corresponds to multiplication by
$G(x)=1+x+x^{n-1}$. We re-derive the invertibility
criterion $3\nmid n$, classify periodic and transient
behaviour, and prove that Hamming weight is non-monotone
for every $n\ge3$. We then factor $x^n-1$ via the
Chinese Remainder Theorem and separate semisimple
contributions governed by finite-field multiplicative
orders from unipotent contributions governed by
Frobenius squaring. We determine the exact local
unipotent orders and obtain a closed-form global order
$r(A)$: $r(A)=2^{m-1}$ for $n=2^m$ with $m\ge2$,
while for $n=2^m q$, $q>1$ odd and $3\nmid n$,
$r(A)=2^m\operatorname{lcm}(m_2,\ldots,m_r)$.
The theory is illustrated by $n=5,7,8,16$ and
computationally verified for small dimensions.
\end{abstract}


\medskip


\bigskip

\section{Introduction}
\label{sec:introduction}

\subsection{Problem Statement}

We study a linear cellular automaton defined on the cyclic graph $C_n$ whose vertices are indexed modulo $n$, each carrying a state in the finite field $\mathbb{F}_2 = \{0,1\}$. The local update rule at each site combines the cell's own state with the states of its two nearest neighbours by addition in $\mathbb{F}_2$ (exclusive-or):
\begin{equation}
    v_i(t+1) = v_{i-1}(t) \oplus v_i(t) \oplus v_{i+1}(t),
\end{equation}
applied synchronously across all sites. In Wolfram's standard numbering of the 256 elementary cellular automaton rules, this update rule is Rule 150, considered here under periodic rather than null boundary conditions. Collecting these local updates into a global linear operator on the configuration space $V_n = \mathbb{F}_2^n$, the global dynamics are governed by
\begin{equation}
    A = I + L + R = I + L + L^{-1},
\end{equation}
where $L$ and $R$ denote the left- and right-shift operators induced by the cyclic adjacency of $C_n$. Under a natural isomorphism between $V_n$ and the quotient ring $\mathbb{F}_2[x]/(x^n - 1)$, the operator $A$ corresponds to multiplication by the polynomial
\begin{equation}
    G(x) = 1 + x + x^{n-1}.
\end{equation}
This correspondence reduces the study of the automaton's long-term behaviour — invertibility, orbit structure, periods, and transients — to questions about the algebraic structure of $G(x)$ inside $\mathbb{F}_2[x] / \langle x^n - 1 \rangle$.

\subsection{Why a Uniform Theory Is Difficult}

Although the local rule is simple, its global dynamics resist a single uniform description for three compounding reasons. First, the state space is discrete and finite rather than continuous, so the classical tools of dynamical systems — Lyapunov functions, continuity arguments, measure-theoretic invariants — do not transfer directly; a natural candidate invariant, the Hamming weight of a configuration, is not monotone under $A$. Second, the underlying algebra is linear only over $\mathbb{F}_2$, where subtraction and addition coincide and many finite-field phenomena (Frobenius structure, characteristic-2 factorization) shape the behaviour in ways that have no continuous analogue. Finally, the ring $\mathbb{F}_2[x] / \langle x^n - 1 \rangle$ does not behave uniformly in $n$: whether $3$ divides $n$ determines whether $A$ is invertible at all, and the factorization of $x^n - 1$ over $\mathbb{F}_2$ mixes separable (semisimple) components with, for even $n$, non-separable components containing nontrivial nilpotents. The result is a system whose qualitative behaviour is intrinsically bimodal — governed by finite-field arithmetic on its semisimple part and by nilpotent, characteristic-2-specific structure on its unipotent part — with, to the best of the author's knowledge, and based on the literature reviewed here, no single existing framework treating both parts uniformly.

\subsection{Mathematical Framework}

This work establishes the following results.

\begin{itemize}
    \item \textbf{Algebraic representation.} Following the standard paradigm for linear cellular automata over finite fields \citep{martin1984}, we represent the evolution operator $A$ on $V_n$ as multiplication by $G(x) = 1 + x + x^{n-1}$ in $\mathbb{F}_2[x] / \langle x^n - 1 \rangle$; this representation is recalled here to make the paper self-contained and to fix the notation used throughout.
    \item \textbf{Invertibility and fixed-point classification.} We re-derive the invertibility criterion $3 \nmid n$ --- equivalent to the known reversibility condition of Nobe and Yura~\citep{nobe2004} --- via the cyclotomic factor $1 + x + x^2$ within our algebraic framework, and use it to classify fixed points and short cycles explicitly for small $n$.
    \item \textbf{Non-monotonicity of the Hamming weight.} We show that the Hamming weight — the most natural candidate for a Lyapunov-type invariant of the dynamics — is not monotone along trajectories of $A$ for any $n \geq 3$, via an explicit pair of configurations whose weight respectively increases and decreases under one application of $A$.
    \item \textbf{Orbit classification.} We show that when $3 \nmid n$ every orbit is purely periodic from the start, and when $3 \mid n$ the state space decomposes into transient trees feeding into periodic attractors; we quantify the resulting information loss and work out the complete orbit structure for $n = 4$.
    \item \textbf{CRT decomposition.} Factoring $x^n - 1$ over $\mathbb{F}_2$, we decompose $\mathbb{F}_2[x] / \langle x^n - 1 \rangle$ via the Chinese Remainder Theorem into local components, reducing global dynamics to independent local dynamics on each factor.
    \item \textbf{Exact local unipotent orders (Theorem~\ref{thm:8-11}).} For every CRT component we give a closed-form, exact determination of the order of its unipotent part, distinguishing the trivial component $(x + 1)$ — which contributes a smaller dyadic exponent than its multiplicity would naively suggest — from the generic case.
    \item \textbf{Global order.} Writing $n = 2^m q$ with $q$ odd, we give the exact order $r(A)$ for every $n$ with $3 \nmid n$: $r(A) = 2^{m-1}$ when $q = 1$, and $r(A) = 2^m \cdot \mathrm{lcm}(m_2, \dots, m_r)$ when $q > 1$, where $m_2, \dots, m_r$ are the multiplicative orders of the semisimple parts on the non-trivial CRT components (Theorem~\ref{thm:8-17}); we prove this value is attained by an orbit.
    \item \textbf{Dyadic structure theorem.} For dimensions $n = 2^m$, we show the dynamics collapse to purely dyadic behaviour, with all orbit periods forced to be powers of two.
\end{itemize}

\subsection{Structure of the Paper}

Section~\ref{sec:related-work} reviews related work on linear cellular automata over finite fields, invertibility criteria, and the connection to finite-field arithmetic. Section~\ref{sec:algebraic} develops the algebraic representation of the automaton as a linear operator on $V_n$ and identifies it with multiplication by a polynomial. Section~\ref{sec:polynomial} constructs the explicit isomorphism $\Phi$ between $V_n$ and $R_n = \mathbb{F}_2[x] / \langle x^n - 1 \rangle$ and proves the Fundamental Dynamical Representation Theorem. Section~\ref{sec:invertibility} establishes the invertibility criterion $3 \nmid n$, describes the kernel of $A$, and quantifies the information loss that occurs when $A$ is not invertible. Section~\ref{sec:orbit} classifies fixed points and periodic orbits, works out the case $n = 4$ in full detail, and proves the non-monotonicity of the Hamming weight. Section~\ref{sec:order} studies the global order $r(A)$ of the evolution operator, its polynomial characterization, and the fully dyadic case $n = 2^m$. Section~\ref{sec:crt} develops the Chinese Remainder decomposition of $R_n$, separates the semisimple and unipotent contributions on each local component via a multiplicative Jordan-type decomposition, establishes the exact local unipotent orders and a dyadic vanishing criterion for the local nilpotent contributions, and derives the exact global order of $A$.
Section~\ref{sec:finite-field} supplies the finite-field background needed to compute local semisimple orders and illustrates the complete theory on four worked examples ($n = 5, 7, 8, 16$). Section~\ref{sec:computational} presents computational experiments verifying the theoretical predictions of Sections ~\ref{sec:invertibility}–~\ref{sec:finite-field}. Section~\ref{sec:conclusions} concludes and lists open directions.

\section{Related Work}
\label{sec:related-work}

\subsection{The Algebraic Paradigm for Linear Cellular Automata}

Linear cellular automata over finite fields have been studied since the 1980s, when Martin, Odlyzko and Wolfram first analysed the dynamics of Wolfram's additive Rule 90 using algebraic tools over $\mathbb{F}_2$, showing that its long-term behaviour could be described through polynomial and matrix methods rather than direct simulation \citep{martin1984}. This established the basic paradigm followed in the present paper: encode a linear automaton's evolution as multiplication by a fixed polynomial in a quotient ring, and read off dynamical properties from the algebraic structure of that ring.

Building on this paradigm, subsequent work obtained closed-form expressions for the minimal and characteristic polynomials of specific linear rules---most notably Rule 90 itself---and used them to give a complete description of its cycle structure, including a proof that the standard basis vectors always generate a cycle of maximal length under periodic boundary conditions \citep{lettieri2006}. Related studies extended the characteristic-polynomial method to cylindrical and higher-dimensional linear automata, and to automata with additional non-local or non-linear terms \citep{stevens1999}.

A separate line of research has focused on invertibility rather than periodicity. Working over general finite commutative rings, Sato gave decidability results and structural criteria for when an additive cellular automaton is globally invertible; later authors sharpened these criteria for specific alphabets, including the prime field case $\mathbb{Z}_p$, and produced explicit algorithms---some of polynomial complexity---for testing reversibility of one-dimensional linear rules with periodic or null boundary conditions \citep{sato1993,cinkir2011,bhattacharjee2016}. The connection between additive cellular automata and finite-field arithmetic more broadly---including global injectivity, the structure of the endomorphism ring of the configuration space, and the role of the Frobenius endomorphism in characteristic $2$---has also been studied directly \citep{vivaldi1994,dow1997}.

Existing work on linear cellular automata over finite commutative rings generally treats reversibility, periodicity and decidability as largely separate questions \citep{cattaneo2004}.

\subsection{Rule 150 in the Literature}

The rule studied in this paper is the operator $G(x) = 1 + x + x^{n-1}$, corresponding to Wolfram's Rule 150, on the cyclic graph $C_n$. This specific rule has a narrower, more directly relevant literature than the general theory surveyed above.

The reversibility of Rule 150 under periodic boundary conditions, for system size $N$ satisfying $N \equiv 1, 2 \pmod 3$, was established by Nobe and Yura~\citep{nobe2004}. This is the same condition---equivalently, $3 \nmid n$---used as the invertibility criterion throughout the present paper (Section~\ref{sec:invertibility}). Hernández Encinas and Martín del Rey~\citep{hernandez2007} went further, giving explicit local transition functions for the inverse cellular automaton in both congruence classes $N \equiv 1, 2 \pmod 3$, rather than only the reversibility criterion itself; the invertibility criterion recovered in Section~\ref{sec:invertibility} of the present paper (Theorem~\ref{thm:5-2}) should accordingly be read as an independent re-derivation of a known result, not a new one.

The periodic structure of Rule 150 was also studied numerically and via eigenvalue analysis by Tadaki~\citep{tadaki1994}, who computed maximum periods and maximum relaxation lengths for finite linear cellular automata under Dirichlet (null) boundary conditions, including explicit period tables for small system sizes. Tadaki's boundary conditions differ structurally from the periodic ones used throughout the present paper---his operator is not a circulant matrix, since it lacks the wraparound coupling between the first and last cells that defines the cyclic graph $C_n$---so a direct numerical match between his period tables and the order formula obtained here (Theorem~\ref{thm:8-17}) is not expected and is not the point of comparison. Instead, Section~\ref{sec:finite-field} discusses where the two settings' qualitative behaviour (e.g. growth of maximal periods with system size) is structurally analogous despite the differing boundary conditions.

Working directly with the matrix representation $A = E + \Lambda + \Lambda^{-1}$, where $\Lambda$ is the cyclic shift matrix---the same representation used in this paper---Kaneko~\citep{kaneko2014} proved, by repeated application of the Frobenius identity $(E + \Lambda + \Lambda^{-1})^{2^m} = E + \Lambda^{2^m} + \Lambda^{-2^m}$, that $A^N = I$ for every system size of the form $N = 2^m(2^k - 1)$; for the specific family $N = 7 \cdot 2^m$ (i.e. $k = 3$), he further showed by direct enumeration that this bound is attained, so the exact multiplicative order of $A$ equals $N$ in that case. For general $k$, his result establishes $A^N = I$ but does not by itself show that $N$ is the minimal such exponent. This result was obtained as a step within a random-matrix-theoretic study of the spacing statistics of Rule 150's periodic orbits, and does not employ a polynomial-ring or Chinese-Remainder-Theorem-based representation: the order is established for this specific family of $N$ by direct matrix manipulation alone.

\subsection{The Gap Addressed by This Paper}

What is missing from the literature surveyed above is a treatment of Rule 150, for arbitrary $n$ with $3 \nmid n$, that:
\begin{enumerate}
    \item[(i)] represents the evolution algebraically via the quotient ring $R_n = \mathbb{F}_2[x]/\langle x^n - 1 \rangle$,
    \item[(ii)] decomposes this ring via the Chinese Remainder Theorem into local components,
    \item[(iii)] within each local component separates the evolution generator into its semisimple and unipotent (nilpotent) parts via a multiplicative Jordan-type decomposition, and
    \item[(iv)] combines these local decompositions into a single closed-form formula for the global order, valid for every admissible $n$ rather than for one special family or one boundary condition.
\end{enumerate}

Kaneko's~\citep{kaneko2014} result is, in this respect, an elementary special case, obtained without factoring $x^n - 1$ and without separating semisimple and nilpotent contributions. Tadaki's~\citep{tadaki1994} results address a related but distinct question---periods under null rather than periodic boundary conditions---via direct eigenvalue computation rather than an algebraic decomposition; the qualitative comparison in Section~\ref{sec:finite-field} is intended to situate this paper relative to his results, not to claim that his results are subsumed by this paper. The general order formula developed in this paper (Theorem~\ref{thm:8-17}, Section~\ref{sec:crt}) is stated for all $n$ with $3 \nmid n$; whether it reduces to Kaneko's result exactly when $q = 2^k - 1$ depends on whether the local semisimple orders $m_i$ satisfy $\operatorname{lcm}(m_2, \dots, m_r) = q$ for that family---a number-theoretic fact that has not been verified in this paper and is not claimed here. If confirmed, it would furnish a consistency check on Theorem~\ref{thm:8-17} against an independently published result; this is left as a direction for Section~\ref{sec:finite-field} and is not presented as an established result in what follows.

The principal novelty of the present paper is therefore not the use of polynomial or matrix methods in themselves---both are established for closely related rules and boundary conditions---but the combination of CRT decomposition, local semisimple--unipotent structure, and a unified global order formula into a single framework, applied to the specific operator $G(x) = 1 + x + x^{n-1}$ under periodic boundary conditions for all $n$ with $3 \nmid n$. The following sections develop this program in a constructive manner.

\section{Algebraic Representation}
\label{sec:algebraic}

Having motivated the problem, we now put it on a precise algebraic footing. The goal of this section is to translate the local, coordinate-wise update rule of the automaton into a single linear operator acting on a finite-dimensional vector space, and then to show that this operator is itself nothing more than multiplication by an explicit polynomial. Once this identification is made, every dynamical question we care about (invertibility, periods, transients) becomes a question in elementary algebra. This translation follows the classical paradigm for linear cellular automata over finite fields established by Martin, Odlyzko, and Wolfram~\citep{martin1984} and developed further by Lettieri, Stevens, and Thomas~\citep{lettieri2006}; the algebraic representation constructed in this section and the next is standard, and is recalled here to fix notation and to make the paper self-contained.

\subsection{The State Space}

\begin{Definition}\label{def:3-1}
Let $\mathbb{F}_2$ denote the finite field with two elements. The global configuration space of the system is the $n$-dimensional vector space
\begin{equation}
V_n = \mathbb{F}_2^n
= \{ v = (v_0, v_1, \dots, v_{n-1}) : v_i \in \mathbb{F}_2 \}.
\end{equation}
\end{Definition}

The coordinates of a configuration $v \in V_n$ are identified with the vertices of the cycle graph $C_n$. Indices are taken modulo $n$ throughout the paper, imposing periodic boundary conditions:
\begin{equation}
    v_{i+n} = v_i, \quad \text{for all } i \in \mathbb{Z}.
\end{equation}
We regard $V_n$ purely as a vector space over $\mathbb{F}_2$; no further structure is assumed at this stage.

\subsection{Fundamental Spatial Operators}

To express the local update rule as a single global operator, we first isolate the elementary spatial moves it is built from.

\begin{Definition}\label{def:3-2}
The left-shift operator $L : V_n \to V_n$ is defined coordinate-wise by $(Lv)_i = v_{i-1}$.
\end{Definition}

\begin{Definition}\label{def:3-3}
The right-shift operator $R : V_n \to V_n$ is defined coordinate-wise by $(Rv)_i = v_{i+1}$.
\end{Definition}

Together with the identity operator $I$, these three maps are linear endomorphisms of $V_n$.

\subsection{The Evolution Operator}

The local rule updates each cell by combining it with its two nearest neighbours via addition in $\mathbb{F}_2$. Packaging this rule into a single operator on all of $V_n$ gives our central object of study.

\begin{Definition}
\label{def:3-4}
The global evolution operator $A : V_n \to V_n$ is
\begin{equation}
    A = L + I + R,
\end{equation}
equivalently, at each coordinate,
\begin{equation}
    (Av)_i = v_{i-1} \oplus v_i \oplus v_{i+1}.
\end{equation}
\end{Definition}

\begin{Proposition}\label{thm:3-5}
For every initial configuration $v_0 \in V_n$, the state at time $t$ satisfies $v_t = A^t v_0$, for all $t \geq 0$.
\end{Proposition}

\begin{proof}
Immediate by induction on $t$, from the recursive definition $v_{t+1} = A v_t$.
\end{proof}

The next step is to understand the algebraic relationship between the two shift operators, since it is this relationship that will let us collapse $A$ to a single polynomial expression.

\begin{Lemma}\label{thm:3-6}
The shift operators satisfy $LR = RL = I$ and $L^n = R^n = I$.
\end{Lemma}

\begin{proof}
Both identities follow directly from the definitions of $L$ and $R$ together with the periodic indexing modulo $n$: composing a left-shift with a right-shift returns every coordinate to its original position, and shifting $n$ times in either direction returns to the identity by periodicity.
\end{proof}

\begin{Proposition}\label{thm:3-7}
The shift operator $L$ generates a cyclic group $\langle L \rangle = \{I, L, L^2, \dots, L^{n-1}\}$ of order $n$, isomorphic to $C_n$.
\end{Proposition}

\begin{proof}
By Lemma~\ref{thm:3-6}, $L^n = I$. If $L^k = I$ for some $k$, then $v_{i-k} = v_i$ for every configuration and every index $i$, which forces $k \equiv 0 \pmod{n}$. Hence $n$ is the least positive integer with $L^n = I$, so $L$ has order exactly $n$.
\end{proof}

We can now state the key identification of this section.

\begin{Lemma}[Key Identity]\label{thm:3-8}
The evolution operator admits the representation
\begin{equation}
    A = L + I + L^{-1}.
\end{equation}
\end{Lemma}

\begin{proof}
By Lemma~\ref{thm:3-6}, $R = L^{-1}$. Substituting into Definition~\ref{def:3-4} gives the result directly.
\end{proof}

\begin{Corollary}\label{thm:3-9}
Since $L^{-1} = L^{n-1}$ by Lemma~\ref{thm:3-6}, the evolution operator is a polynomial in $L$:
\begin{equation}
    A = I + L + L^{n-1}.
\end{equation}
\end{Corollary}

Because $A$ is expressed purely as a polynomial evaluated at the single generator $L$ of a cyclic group of order $n$, the entire dynamical system can be re-interpreted algebraically: $L$ behaves exactly like multiplication by the indeterminate $x$ inside the quotient ring $\mathbb{F}_2[x] / \langle x^n - 1 \rangle$, and consequently
\begin{equation}
    A \leftrightarrow G(x) = 1 + x + x^{n-1}
\end{equation}
as an element of that ring. This polynomial coincides with the transition polynomial used for Rule 150 in Hernández Encinas and Martín del Rey~\citep{hernandez2007} and Nobe and Yura~\citep{nobe2004}; the reformulation $A \leftrightarrow G(x)$ given here is the same identification, expressed in the quotient-ring language of Section~\ref{sec:polynomial}. From this point on, the dynamics of the automaton — its periods, its invertibility, its transient structure — are governed entirely by the algebraic behaviour of the single polynomial $G(x) = 1 + x + x^{n-1}$ inside $\mathbb{F}_2[x] / \langle x^n - 1 \rangle$: its factorization, its greatest common divisor with $x^n - 1$, and the multiplicative orders of its irreducible factors. Section~\ref{sec:polynomial} makes this correspondence precise by constructing an explicit isomorphism $\Phi$ between $V_n$ and $\mathbb{F}_2[x] / \langle x^n - 1 \rangle$ under which $A$ corresponds exactly to multiplication by $G(x)$.

\section{Polynomial Model}
\label{sec:polynomial}

This section constructs an explicit isomorphism between $V_n$ and a quotient ring of polynomials, and verifies that under this isomorphism the evolution operator $A$ corresponds to multiplication by $G(x)$. The construction of $\Phi$ and the resulting identity $\Phi(A\mathbf{v}) = G(x) \cdot \Phi(\mathbf{v})$ (Theorem~\ref{thm:4-7}) are the same reformulation used throughout the linear-cellular-automata literature cited in Section~\ref{sec:related-work}; the isomorphism itself is standard, and what follows in Sections~\ref{sec:invertibility}--\ref{sec:finite-field} is the analysis carried out within it.

\subsection{The Quotient Ring $R_n$}

\begin{Definition}\label{def:4-1}
Let $\mathbb{F}_2[x]$ be the polynomial ring over $\mathbb{F}_2$. Define
\begin{equation}
    R_n = \mathbb{F}_2[x] / \langle x^n - 1 \rangle.
\end{equation}
\end{Definition}

The defining relation of $R_n$ is $x^n \equiv 1 \pmod{x^n - 1}$, which is the algebraic analogue of $L^n = I$ established in Section~\ref{sec:algebraic}.

\subsection{The Isomorphism $\Phi$}

\begin{Definition}\label{def:4-2}
The map $\Phi : V_n \to R_n$ is defined by
\begin{equation}
    \Phi(v_0, v_1, \dots, v_{n-1}) = v_0 + v_1 x + v_2 x^2 + \dots + v_{n-1} x^{n-1}.
\end{equation}
\end{Definition}

\begin{Proposition}\label{thm:4-3}
$\Phi$ is an isomorphism of vector spaces over $\mathbb{F}_2$.
\end{Proposition}

\begin{proof}
Let $\{e_0, \dots, e_{n-1}\}$ be the standard basis of $V_n$. By Definition~\ref{def:4-2}, $\Phi(e_i) = x^i$ for each $i$. Since $\{1, x, \dots, x^{n-1}\}$ is the standard basis of $R_n$, $\Phi$ maps a basis of $V_n$ bijectively onto a basis of $R_n$. It is $\mathbb{F}_2$-linear by construction. Hence $\Phi$ is a vector space isomorphism.
\end{proof}

\subsection{Shift Operators Under $\Phi$}

\begin{Theorem}\label{thm:4-4}
For all $v \in V_n$, $\Phi(Lv) = x \cdot \Phi(v)$.
\end{Theorem}

\begin{proof}
Write $v = (v_0, \dots, v_{n-1})$, so $Lv = (v_{n-1}, v_0, \dots, v_{n-2})$. Then
\begin{equation}
    \Phi(Lv) = v_{n-1} + v_0 x + v_1 x^2 + \dots + v_{n-2} x^{n-1}.
\end{equation}
On the other hand,
\begin{equation}
    x \cdot \Phi(v) = v_0 x + v_1 x^2 + \dots + v_{n-2} x^{n-1} + v_{n-1} x^n.
\end{equation}
Since $x^n \equiv 1 \pmod{x^n - 1}$, the term $v_{n-1} x^n$ reduces to $v_{n-1}$, and the two expressions coincide.
\end{proof}

\begin{Corollary}\label{thm:4-5}
For every $k \geq 0$, $\Phi(L^k v) = x^k \cdot \Phi(v)$. This follows by repeated application of Theorem~\ref{thm:4-4}.
\end{Corollary}

\subsection{Polynomial Form of the Evolution Operator}

By Lemma~\ref{thm:3-8}, $A = L + I + L^{-1}$. Applying $\Phi$ and using its linearity:
\begin{equation}
    \Phi(Av) = \Phi(Lv) + \Phi(v) + \Phi(L^{-1}v).
\end{equation}
Since $L^{-1} = L^{n-1}$, Theorem~\ref{thm:4-4} and Corollary~\ref{thm:4-5} give $\Phi(Lv) = x \cdot \Phi(v)$ and $\Phi(L^{-1}v) = x^{n-1} \cdot \Phi(v)$. Substituting:
\begin{equation}
    \Phi(Av) = (1 + x + x^{n-1}) \cdot \Phi(v).
\end{equation}

\begin{Definition}\label{def:4-6}
The generator polynomial of the system is $G(x) = 1 + x + x^{n-1} \in R_n$. By the calculation above,
\begin{equation}
    \Phi(Av) = G(x) \cdot \Phi(v).
\end{equation}
\end{Definition}

The action of $A$ on $V_n$ corresponds, under $\Phi$, to multiplication by $G(x)$ on $R_n$.

\subsection{The Fundamental Dynamical Representation Theorem}

Let $v_0 \in V_n$ be an initial configuration, and let $P_0(x) = \Phi(v_0)$.

\begin{Theorem}\label{thm:4-7}
For every $t \geq 0$, the polynomial representation $P_t(x) = \Phi(v_t)$ of $v_t = A^t v_0$ satisfies
\begin{equation}
    P_t(x) = P_0(x) \cdot G(x)^t.
\end{equation}
\end{Theorem}

\begin{proof}
By induction on $t$, using $\Phi(Av) = G(x)\Phi(v)$ at each step, $\Phi(A^t v_0) = G(x)^t \Phi(v_0)$, since $\Phi$ is linear and $A^t$ corresponds to $t$ successive multiplications by $G(x)$. Hence
\begin{equation}
    P_t(x) = \Phi(v_t) = \Phi(A^t v_0) = G(x)^t \Phi(v_0) = P_0(x) G(x)^t.
\end{equation}
\end{proof}

\subsection{Consequences}

Theorem~\ref{thm:4-7} replaces the coordinate-wise update rule $v_i(t+1) = v_{i-1}(t) \oplus v_i(t) \oplus v_{i+1}(t)$ with a single algebraic relation, $P_t(x) = P_0(x) G(x)^t$, in the ring $R_n$. Three consequences follow directly.

\begin{enumerate}
    \item Time evolution corresponds to exponentiation of $G(x)$.
    \item Long-term behaviour — periods, invertibility — reduces to algebraic properties of $G(x)$: its factorization, its gcd with $x^n - 1$, and the multiplicative orders of its irreducible factors.
    \item If the trajectory of $v_0$ is periodic, its period is the least $k \geq 1$ such that $(G(x)^k - 1) P_0(x) \equiv 0 \pmod{x^n - 1}$. The multiplicative order of $G(x)$ in $R_n$, when defined, bounds all such periods from above; the period of a specific $v_0$ depends additionally on $P_0(x)$. This is developed in Sections ~\ref{sec:order}-~\ref{sec:finite-field}.
\end{enumerate}

\section{Invertibility}
\label{sec:invertibility}

The polynomial representation established in Section~\ref{sec:polynomial} immediately reduces invertibility of the evolution operator to invertibility of its generator polynomial. Indeed, by Theorem~\ref{thm:4-7}, $A$ corresponds under $\Phi$ to the multiplication map $P(x) \mapsto G(x)P(x)$ on $R_n$. Hence $A$ is invertible if and only if $G(x)$ is a unit of $R_n$. The invertibility criterion obtained below (Theorem~\ref{thm:5-2}) is not new: it is equivalent to the condition $n \equiv 1, 2 \pmod 3$ established by Nobe and Yura~\citep{nobe2004} for reversibility of Rule 150 under periodic boundary conditions. The derivation given here, via the cyclotomic polynomial $\Phi_3(x) = 1+x+x^2$, is presented as an independent, self-contained re-derivation within the algebraic framework of this paper, rather than as a new result.

\subsection{Criterion for Invertibility}

\begin{Theorem}\label{thm:5-1}
$A$ is invertible if and only if $\gcd(G(x), x^n - 1) = 1$ in $\mathbb{F}_2[x]$.
\end{Theorem}

\begin{proof}
Multiplication by $G(x)$ is invertible in $\mathbb{F}_2[x]/\langle x^n - 1 \rangle$ if and only if $G(x)$ is a unit of the quotient ring, which by Bézout's identity holds if and only if $\gcd(G(x), x^n - 1) = 1$. Since $\Phi$ is an isomorphism, $A$ is invertible under exactly this condition.
\end{proof}

\subsection{Reduction of the GCD Condition}

Since $x^n \equiv 1 \pmod{x^n - 1}$,
\begin{equation}
    xG(x) = x + x^2 + x^n \equiv 1 + x + x^2 \pmod{x^n - 1}.
\end{equation}
As $x$ is a unit of $\mathbb{F}_2[x]$ coprime to $x^n - 1$, this gives
\begin{equation}
    \gcd(G(x), x^n - 1) = \gcd(1 + x + x^2, x^n - 1).
\end{equation}
The polynomial $\Phi_3(x) = 1 + x + x^2$ is the third cyclotomic polynomial reduced mod $2$. Invertibility of $A$ therefore reduces to the question of whether $\Phi_3(x)$ divides $x^n - 1$.

\begin{Theorem}\label{thm:5-2}
$A$ is non-invertible if and only if $3 \mid n$.
\end{Theorem}

\begin{proof}
$1 + x + x^2$ is irreducible over $\mathbb{F}_2$, so $\gcd(1+x+x^2, x^n-1) \neq 1$ iff $1+x+x^2 \mid x^n - 1$. Let $\omega$ be a root of $1+x+x^2$ in an extension of $\mathbb{F}_2$. Then $1+\omega+\omega^2 = 0$, so $\omega^3 = 1$ and $\omega \neq 1$; $\omega$ has multiplicative order $3$. The divisibility $1+x+x^2 \mid x^n-1$ holds iff $\omega$ is a root of $x^n - 1$, i.e.\ $\omega^n = 1$, which holds iff $3 \mid n$. Combined with Theorem~\ref{thm:5-1}, this gives the claim.
\end{proof}

\begin{Remark}
Since $3 \mid n$ is equivalent to $n \not\equiv 1, 2 \pmod 3$, Theorem~\ref{thm:5-2} recovers exactly the reversibility criterion of Nobe and Yura~\citep{nobe2004}. Hernández Encinas and Martín del Rey~\citep{hernandez2007} go further and compute the explicit local transition functions of the inverse cellular automaton for both residue classes $n \equiv 1, 2 \pmod 3$; the present section does not construct this inverse explicitly, and is concerned only with re-deriving the criterion itself as a consequence of the algebraic framework of Section~\ref{sec:algebraic}.
\end{Remark}

\begin{Corollary}\label{thm:5-3}
$A$ preserves complete information if and only if $3 \nmid n$. If $3 \mid n$, $A$ has a non-trivial kernel and distinct configurations can share a successor.
\end{Corollary}

\subsection{The Kernel of $A$}

\begin{Definition}\label{def:5-4}

$\ker(A) = \{ v \in V_n : Av = 0 \}$. Under $\Phi$,
$\ker(A)$ corresponds to ${ P(x) \in R_n : G(x)P(x) \equiv 0 \pmod{x^n - 1}
}$.

\end{Definition}

\begin{Theorem}\label{thm:5-5}

If $3 \nmid n$, then $\ker(A)={0}$. If $3\mid n$, then

$$
\dim_{\mathbb{F}_2}\ker(A)=2.
$$

In particular, $|\ker(A)|=4$.

\end{Theorem}

\begin{proof}

Let

$$
D(x)=\gcd(G(x),x^n-1).
$$

For a multiplication map by $G(x)$ on
$R_n=\mathbb{F}_2[x]/\langle x^n-1\rangle$, the dimension of the kernel is
equal to $\deg D(x)$. Indeed, write

$$
G(x)=D(x)G_1(x),
\qquad
x^n-1=D(x)H_1(x),
$$

where $\gcd(G_1(x),H_1(x))=1$. Then

$$
G(x)P(x)\equiv 0 \pmod{x^n-1}
$$

if and only if

$$
D(x)H_1(x)\mid D(x)G_1(x)P(x),
$$

which is equivalent to

$$
H_1(x)\mid G_1(x)P(x).
$$

Since $\gcd(G_1(x),H_1(x))=1$, this holds if and only if
$H_1(x)\mid P(x)$. Hence every element of the kernel has a representative of
the form

$$
P(x)=H_1(x)Q(x),
$$

with $Q(x)$ determined modulo $D(x)$. Therefore the kernel is
isomorphic as an $\mathbb{F}_2$-vector space to
$\mathbb{F}_2[x]/\langle D(x)\rangle$, and consequently

$$
\dim_{\mathbb{F}_2}\ker(A)=\deg D(x).
$$

If $3 \nmid n$, Theorem~\ref{thm:5-2} gives
$D(x)=1$, so $\ker(A)={0}$. If $3\mid n$, then
Theorem~\ref{thm:5-2} and the GCD reduction above give

$$
D(x)=1+x+x^2,
$$

and therefore

$$
\dim_{\mathbb{F}_2}\ker(A)=\deg(1+x+x^2)=2.
$$

Since a two-dimensional vector space over $\mathbb{F}_2$ has four elements,
$|\ker(A)|=4$.

\end{proof}

\begin{Example}\label{ex:5-6}

$n = 3$. Here $3 \mid 3$, so $A$ is non-invertible by
Theorem~\ref{thm:5-2}. Explicitly,

\begin{equation}
    A(v_0,v_1,v_2)
    =
    (v_0+v_1+v_2,\,
     v_0+v_1+v_2,\,
     v_0+v_1+v_2).
\end{equation}

so $(v_0,v_1,v_2) \in \ker(A)$ iff $v_0+v_1+v_2 = 0$. Hence

\begin{equation}
    \ker(A)
    =
    \left\{
        (0,0,0),\,
        (0,1,1),\,
        (1,0,1),\,
        (1,1,0)
    \right\}.
\end{equation}

a two-dimensional subspace of $V_3$. In particular $A(1,0,1)
= (0,0,0)$: a non-zero configuration maps to the zero state.

\end{Example}

\subsection{Information Loss}

\begin{Proposition}\label{thm:5-7}

For $v, w \in V_n$, $Av = Aw$ if and only if $v + w \in
\ker(A)$.

\end{Proposition}

\begin{proof}

$Av = Aw$ iff $Av + Aw = 0$ iff $A(v+w) = 0$ (linearity over
$\mathbb{F}_2$) iff $v + w \in \ker(A)$.

\end{proof}

Two configurations share a successor exactly when they lie
in the same coset of $\ker(A)$. Writing $d=\dim(\ker(A))$, every state in
$\operatorname{im}(A)$ has exactly $2^d$ predecessors under one application
of $A$. Thus, one application of $A$ identifies $2^d$ configurations in each
fiber of the map $A:V_n\to\operatorname{im}(A)$. In particular, when
$3\mid n$, we have $d=2$, so every reachable state has exactly four immediate
predecessors.

\subsection{Reversibility of the Dynamics}

\begin{Lemma}\label{thm:5-8}
Every self-map $f : X \to X$ of a finite set $X$ decomposes each connected component of its functional graph into a single cycle together with rooted trees feeding into it; if $f$ is injective, no such trees occur, and $f$ is a permutation of $X$.
\end{Lemma}

\begin{proof}
Standard fact about functional graphs of finite sets: since $X$ is finite, every forward orbit under $f$ is eventually periodic, so each component contains exactly one cycle, with the remaining vertices organized into trees rooted on that cycle. If $f$ is injective, it is a bijection of the finite set $X$, hence a permutation, and every permutation decomposes into disjoint cycles with no transient vertices.
\end{proof}

\begin{Theorem}\label{thm:5-9}
If $3 \nmid n$, every configuration lies on a periodic orbit; $V_n$ contains no transient states.
\end{Theorem}

\begin{proof}
By Theorem~\ref{thm:5-2}, $A$ is invertible, hence injective, and $V_n$ is finite. The claim follows from Lemma~\ref{thm:5-8}.
\end{proof}

\begin{Theorem}\label{thm:5-10}
If $3 \mid n$, the state-space digraph contains transient trees feeding into periodic attractors.
\end{Theorem}

\begin{proof}
By Theorem~\ref{thm:5-2}, $A$ is not invertible, hence not injective (Proposition~\ref{thm:5-7}), and $V_n$ is finite. Applying Lemma~\ref{thm:5-8} to $A$ gives the claim.
\end{proof}

\subsection{Summary}

The divisibility $3 \mid n$ determines the qualitative
structure of the dynamics. If $3 \nmid n$, $A$ is invertible and $V_n$
decomposes into disjoint periodic cycles. If $3 \mid n$, $\ker(A)$ is
non-trivial, in fact

$$
\dim_{\mathbb{F}_2}\ker(A)=2
\qquad\text{and}\qquad
|\ker(A)|=4,
$$

so every reachable state has exactly four immediate predecessors. Moreover,
trajectories merge, and the orbit graph contains transient trees
feeding into periodic cycles. Thus, the arithmetic condition $3 \mid n$
completely determines whether the dynamics are reversible or irreversible.

\section{Orbit Structure}

\label{sec:orbit}

\subsection{Definitions and Finiteness of Orbits}

\begin{Definition}\label{def:6-1}

The orbit of $v \in V_n$ is $O(v) = \{ A^t v : t \geq 0 \}$.

\end{Definition}

\begin{Lemma}\label{thm:6-2}

For every $v \in V_n$, $O(v)$ is finite and $|O(v)| \leq
2^n$.

\end{Lemma}

\begin{proof}

$V\_n$ is finite with $2^n$ elements, so the sequence $v, Av,
A^2v, \ldots$ must eventually repeat: there exist $t_0 \geq 0$ and $L \geq 1$
with $A^{t_0+L}v = A^{t_0}v$, hence $O(v)$ is finite.

\end{proof}

\begin{Corollary}\label{thm:6-3}

Every trajectory decomposes into a transient part followed
by a periodic cycle:

$$
v_0 \to v_1 \to \cdots \to v_{t_0-1} \to c_0 \to \cdots \to
c_{L-1} \to c_0.
$$

\end{Corollary}

\begin{Theorem}\label{thm:6-4}

If $3 \nmid n$, $A$ is invertible and $V_n$ decomposes into
disjoint periodic cycles; every orbit is purely periodic ($t_0 = 0$).

\end{Theorem}

\begin{proof}

By Theorem~\ref{thm:5-2}, $A$ is invertible, hence a
bijection of the finite set $V_n$, hence a permutation of $V_n$.
Permutations decompose into disjoint cycles, so no transients occur.

\end{proof}

\subsection{Smallest Dimension with Nontrivial Periodic Orbits}

\begin{Definition}\label{def:6-5}

$v \in V_n$ is a fixed point if $Av = v$.

\end{Definition}

\begin{Definition}\label{def:6-6}

A periodic orbit has minimal period $L$ if $A^L\mathbf{v} =
\mathbf{v}$ and $A^k\mathbf{v} \neq \mathbf{v}$ for $1 \le k < L$. An orbit
is called nontrivial if $L > 1$, and trivial (a fixed point) if $L = 1$.

\end{Definition}

For $n = 1, 2$, $A$ is the identity, so every configuration
is a fixed point. For $n = 3$, $A$ is non-invertible (Theorem~\ref{thm:5-2})
and every trajectory collapses to a fixed point in at most one step.

\begin{Theorem}\label{thm:6-7}

The smallest $n$ admitting a nontrivial periodic orbit is $n = 4$.

\end{Theorem}

\begin{proof}

We rule out $n = 1, 2, 3$ directly and exhibit a nontrivial
periodic orbit at $n = 4$.

\textbf{Case $n = 1$:} $G(x) = 1 + x + x^0 = 1 + x + 1 = x$,
and $x \equiv 1 \pmod{x - 1}$, so $G(x) \equiv 1$ and $A = I$. Every
configuration is a fixed point.

\textbf{Case $n = 2$:} $G(x)=1+x+x=1$ in
$\mathbb{F}_2[x] / \langle x^n - 1 \rangle$, so again $A=I$ and every configuration is a fixed point.

\textbf{Case $n = 3$:} By Example~\ref{ex:5-6},
$A(v_0,v_1,v_2) = (s,s,s)$ with $s = v_0+v_1+v_2$, so
$\operatorname{Im}(A) = \{(0,0,0), (1,1,1)\}$. Both elements of
$\operatorname{Im}(A)$ are themselves fixed points, since applying $A$ to
either gives $s' = 1+1+1 = 1$, unchanged. By Corollary~\ref{thm:6-3},
every trajectory eventually enters $\operatorname{Im}(A)$ and remains there;
since $\operatorname{Im}(A)$ contains only fixed points, no periodic orbit of
length greater than $1$ exists for $n = 3$.

\textbf{Case $n = 4$:} Example~\ref{ex:6-11} below exhibits
the explicit orbit $(1,0,0,0) \leftrightarrow (1,1,0,1)$ of length $2$.

Combining the four cases, $n = 4$ is exactly the smallest
dimension admitting a nontrivial periodic orbit.

\end{proof}

\subsection{The Case $n = 4$}
\label{subsec:n4}
\begin{Theorem}\label{thm:6-8}

For $n = 4$, every orbit has period $L \in \{1, 2\}$.

\end{Theorem}

\begin{proof}

Here $G(x) = 1 + x + x^3$. Over $\mathbb{F}_2$,

$$
G(x)^2 = 1 + x^2 + x^6.
$$

Reducing modulo $x^4 - 1$ gives $x^6 \equiv x^2$, hence

$$
G(x)^2 \equiv 1.
$$

Therefore, $A^2 = I$, and every orbit has period dividing $2$.

\end{proof}

\begin{Proposition}[Classification of fixed
points]\label{thm:6-9}

Solving $Av = v$ is equivalent to $(A+I)v=0$, or in
polynomial form $(G(x)-1)P(x) \equiv 0 \pmod{x^4-1}$. With $G(x) = 1+x+x^3$,
this becomes $(x+x^3)P(x) \equiv 0$. Over $\mathbb{F}_2$,

$$
x + x^3 = x(1+x)^2, \qquad x^4 - 1 = (1+x)^4.
$$

Since $x$ is a unit, the condition reduces to $(1+x)^2P(x)
\equiv 0 \pmod{(1+x)^4}$, so $P(x)$ must be divisible by $(1+x)^2$. The
solution space is $P(x) = (C_0+C_1x)(1+x^2)$, a $2$-dimensional subspace of
$V_4$ with $2^2 = 4$ elements.

\end{Proposition}

\begin{Proposition}[Two-cycles]\label{thm:6-10}

By Theorem~\ref{thm:6-8}, $A^2 = I$, so every orbit has
length at most $2$, and by Theorem~\ref{thm:6-4} ($3 \nmid 4$) there are no
transients. The fixed-point subspace has $4$ elements, leaving $16-4 = 12$
configurations in cycles of length $2$, giving $12/2 = 6$ nontrivial periodic
orbits.

\end{Proposition}

\begin{table}[H]

\caption{Orbit counts for the Rule 150 automaton with
$n=4$.\label{tab:n4-orbits}}

\centering

\begin{tabularx}{\textwidth}{lCCCr}

\toprule

\textbf{Orbit type} & \textbf{Period ($L$)} &
\textbf{Characterization} & 

\textbf{Number of orbits} & \textbf{Total
states} \\\\

\midrule

Fixed points & 1 & $v \in \ker(A \oplus I)$ & 4
& 4 \\\\

Nontrivial periodic orbits & 2 & $v \notin \ker(A
\oplus I)$ & 6 & 12 \\\\

\bottomrule

\end{tabularx}

\end{table}

\begin{Example}\label{ex:6-11}

Take $v = (1,0,0,0)$, i.e.\ $P(x) = 1$. Then $Av
\leftrightarrow G(x)\cdot 1 = 1+x+x^3$, i.e.\ $(1,1,0,1) \neq v$. Applying $A$
again, $A^2v \leftrightarrow G(x)^2 \equiv 1$, so $A^2v = v$. Hence $(1,0,0,0)
\leftrightarrow (1,1,0,1)$ is a nontrivial periodic orbit of length $2$.

\end{Example}

\subsection{Dynamical Consequences}

The relation $G(x)^2 \equiv 1 \pmod{x^4-1}$ gives $A^2=I$, so the
order of $A$ divides $2$. Since Example~\ref{ex:6-11} exhibits a non-fixed
configuration, $A\neq I$, and therefore $\operatorname{ord}(A)=2$.

More generally, when $3\nmid n$, let $r(A)$ denote the multiplicative order
of $G(x)$ in $R_n$. Then every orbit is periodic with period dividing $r(A)$,
and nontrivial periodic orbits exist exactly when $r(A)>1$.

This order is studied systematically in
Section~\ref{sec:order} .

Beyond the classification of periodic orbits, it is natural
to ask whether some simple statistic of a configuration could serve as a
Lyapunov function for the dynamics. We show that this fails already for the
most natural candidate, the Hamming weight.

\subsection{Non-Monotonicity of the Hamming Weight}

\begin{Definition}\label{def:6-12}

The Hamming weight of $v \in V_n$ is $\operatorname{wt}(v) =
|\{i : v_i \neq 0\}|$.

\end{Definition}

\begin{Theorem}\label{thm:6-13}

For every $n \geq 3$, the Hamming weight is neither nonincreasing nor
nondecreasing under $A$: there exist configurations $v, v' \in V_n$ with

$$
\operatorname{wt}(Av) > \operatorname{wt}(v)
\quad\text{and}\quad
\operatorname{wt}(Av') < \operatorname{wt}(v').
$$

In particular, neither $\operatorname{wt}(Av) \leq \operatorname{wt}(v)$
nor $\operatorname{wt}(Av) \geq \operatorname{wt}(v)$
holds for all $v \in V_n$.

\end{Theorem}

\begin{proof}

Let $J = (1,1,\ldots,1)$ denote the all-ones configuration.
Since $(AJ)_i = 1+1+1 = 1$ for every $i$, we have $AJ = J$.

Take $v = e\_0$, the configuration with a single $1$ in
coordinate $0$. Then $(Av)_i = v_{i-1}+v_i+v_{i+1}$ is nonzero exactly at $i
\in \{-1,0,1\} \pmod{n}$, three pairwise distinct indices since $n \geq 3$.
Hence $\operatorname{wt}(v) = 1$ while $\operatorname{wt}(Av) = 3 >
\operatorname{wt}(v)$.
Take $v' = J + e_0$ (addition in $\mathbb{F}_2$), so
$\operatorname{wt}(v') = n-1$. By linearity of $A$ and $AJ = J$,

$$
Av' = AJ + Ae_0 = J + Ae_0.
$$

Since $Ae_0$ has exactly the three nonzero coordinates
$-1, 0, 1\pmod{n}$ computed above, adding it to $J$ flips those three
coordinates of $J$ from $1$ to $0$ and leaves the rest at $1$. Hence
$\operatorname{wt}(Av') = n-3 < n-1 = \operatorname{wt}(v')$.

\end{proof}

Thus the Hamming weight cannot serve as a Lyapunov-type
function for these dynamics. Periodicity and orbit structure are instead
determined directly from the arithmetic of $R_n = \mathbb{F}_2[x] / \langle x^n - 1 \rangle$,
beginning with the operator order computed in the next section.

\section{Order of the Evolution Operator}

\label{sec:order}

\subsection{Global Order}

\begin{Definition}\label{def:7-1}

The order of the evolution operator of $A$, denoted $r(A)$,
is the smallest positive integer $m$ such that $A^m = I$.

\end{Definition}

\begin{Proposition}\label{thm:7-2}

$r(A)$ exists if and only if $A$ is invertible, i.e.\ by
Theorem~\ref{thm:5-2}, if and only if $3 \nmid n$.

\end{Proposition}

\begin{proof}

If $3 \mid n$, then $\ker(A) \neq \{0\}$, so $A$ is not
injective. If $A$ is not injective, then neither is $A^m$ for any $m \geq 1$,
whereas the identity operator is injective. Hence no power of $A$ can equal
$I$. Therefore the order of $A$ is undefined.

Conversely, suppose that $A$ is invertible. By Theorem~\ref{thm:5-9}, $A$ is a
permutation of the finite set $V_n$. Every permutation of a finite set has
finite order, so some positive power of $A$ equals $I$. Thus $r(A)$ exists.

\end{proof}

The order of the evolution operator provides a universal
upper bound for all orbit lengths and therefore serves as the principal global
invariant of the dynamics.

\subsection{Orbit Periods and Global Order}

\begin{Definition}\label{def:7-3}

For a periodic $v \in V_n$, the minimal period is $L(v) =
\min\{ k \in \mathbb{N} : A^k v = v \}$.

\end{Definition}

$L(v)$ depends on the orbit; $r(A)$ is a global invariant
with $A^{r(A)} v = v$ for all $v \in V_n$.

\begin{Example}\label{ex:7-4}

For $n = 4$, $A^2 = I$, so $r(A) = 2$. Fixed points have
$L(v) = 1$; non-trivial orbits have $L(v) = 2$. All local periods divide the
global order.

\end{Example}

\begin{Theorem}\label{thm:7-5}

For every periodic $v \in V_n$, $L(v) \mid r(A)$.

\end{Theorem}

\begin{proof}

Let $r = r(A)$, so $A^r = I$ and $A^r v = v$. Write $r =
qL(v) + \rho$ with $0 \leq \rho < L(v)$. Then $v = A^r v = A^\rho v$. By
minimality of $L(v)$, $\rho = 0$, so $L(v) \mid r$.

\end{proof}

\subsection{Polynomial Characterization}

\begin{Theorem}\label{thm:7-6}

$r(A)$ equals the multiplicative order of $G(x)$ in $R_n$:

$$
r(A) = \min\{ m \in \mathbb{N} : G(x)^m \equiv 1 \pmod{x^n - 1} \}.
$$

\end{Theorem}

\begin{proof}

Under $\Phi$, $A^m$ corresponds to multiplication by
$G(x)^m$. $A^m = I$ holds iff $G(x)^m P(x) \equiv P(x)$ for all $P(x) \in R_n$,
iff $G(x)^m \equiv 1$.

\end{proof}

This reduces the study of orbit periods to computing the
multiplicative order of $G(x)$ in $R_n$.

\begin{Corollary}\label{thm:7-7}

Every periodic orbit length divides $r(A)$ and is therefore bounded by
$r(A)$.

\end{Corollary}

\begin{Corollary}\label{thm:7-8}

If $r(A) = p$ is prime, every periodic configuration has
period $L(v) \in \{1, p\}$.

\end{Corollary}

\subsection{Dyadic Dimensions}

Consider $n = 2^m$. The algebra simplifies via the Frobenius
endomorphism.

\begin{Lemma}\label{thm:7-9}

For $P(x) \in \mathbb{F}_2[x]$ and $k \geq 0$, $P(x)^{2^k} =
P(x^{2^k})$.

\end{Lemma}

\begin{proof}

Follows from $(a+b)^2 = a^2 + b^2$ in characteristic $2$,
applied iteratively.

\end{proof}

\begin{Theorem}\label{thm:7-10}

If $n = 2^m$, then $A^n = I$.

\end{Theorem}

\begin{proof}

It suffices to show $G(x)^n \equiv 1 \pmod{x^n - 1}$. By
Lemma~\ref{thm:7-9}, $G(x)^n = 1 + x^n + x^{n(n-1)}$. Since $x^n \equiv 1$,
$x^{n(n-1)} = (x^n)^{n-1} \equiv 1$, so $G(x)^n \equiv 1+1+1 = 1$. Hence $A^n =
I$.

\end{proof}

\begin{Remark}

Theorem~\ref{thm:7-10} is the special case $q = 1$ of the
more general identity $A^{2^m(2^k-1)} = I$ proved by Kaneko (2014, Theorem 3.1)
for all $m, k$ with $2^m(2^k-1) \equiv 1, 2 \pmod 3$; the elementary proof
given here, via Lemma~\ref{thm:7-9} alone, suffices for the pure power-of-two
case and is included because it is used directly in Section~\ref{sec:crt}.

\end{Remark}

\begin{Corollary}\label{thm:7-11}

If $n = 2^m$, $r(A) \mid n$.

\end{Corollary}

\begin{Corollary}\label{thm:7-12}

For $n = 2^m$, every configuration is periodic and has period
$L(v)=2^k$ for some $0\le k\le m$.

\end{Corollary}

\begin{proof}

$L(v) \mid r(A) \mid 2^m$, so $L(v)$ is a power of two at
most $2^m$.

\end{proof}

\begin{Corollary}\label{thm:7-13}

Only powers of two occur as orbit lengths when $n$ is a
power of two.

\end{Corollary}

\subsection{Transition to the General Case}

For $n = 2^m$, all orbit lengths are powers of two and $r(A)
\mid n$. For general $n$, the structure of $x^n-1$ depends not only on its
factorization into irreducible polynomials over $\mathbb{F}_2$, but also on the
multiplicities of those factors. Accordingly, the order of $G(x)$ must be
analysed on the resulting local components
$\mathbb{F}_2[x]/\langle f_i(x)^{e_i} \rangle$. This is developed in
Section~\ref{sec:crt} via the Chinese Remainder Theorem.

\section{CRT Decomposition}

\label{sec:crt}

\subsection{Separability of $x^n - 1$}

\begin{Proposition}\label{thm:8-1}

If $n$ is odd, $x^n - 1$ is separable over $\mathbb{F}_2$.

\end{Proposition}

\begin{proof}

Let $f(x) = x^n - 1$. Since $n$ is odd, $n \equiv 1 \pmod{2}$, so $f'(x) = n
\cdot x^{n-1} = x^{n-1}$, and $\gcd(x^n-1, x^{n-1}) = 1$, so $f$ is separable.

\end{proof}

\begin{Proposition}\label{thm:8-2}

If $n = 2^m \cdot q$ with $q$ odd, then $x^n - 1 = (x^q - 1)^{2^m}$; in
particular $x^n - 1$ is not separable when $m \geq 1$.

\end{Proposition}

\begin{proof}

Since $x^{2^m q} - 1 = (x^q)^{2^m} - 1 = (x^q - 1)^{2^m}$, the identity follows
by repeated application of the Frobenius identity $(a+b)^2 = a^2 + b^2$ in
characteristic $2$.

\end{proof}

\begin{Corollary}\label{thm:8-3}

$x^n - 1$ is separable if and only if $n$ is odd.

\end{Corollary}

\subsection{CRT Decomposition}

Let $x^n - 1 = f_1(x)^{e_1} \cdots f_r(x)^{e_r}$ be the factorization into
distinct irreducible polynomials over $\mathbb{F}_2$.

\begin{Theorem}[CRT Decomposition]\label{thm:8-4}

There is a ring isomorphism

\[
R_n \cong \prod_{i=1}^{r} \mathbb{F}_2[x]/\langle f_i(x)^{e_i} \rangle.
\]

\end{Theorem}

\begin{proof}

The factors $f_i(x)^{e_i}$ are pairwise coprime; the isomorphism is the Chinese
Remainder Theorem.

\end{proof}

Under this isomorphism, a configuration's polynomial $P(x)$ splits into
components $(P_1(x),\ldots,P_r(x))$, and $A$ acts componentwise on each CRT
component, by multiplication by $G(x)$ modulo $f_i(x)^{e_i}$. The system
decomposes into $r$ independent subsystems.

\subsection{Local Orders}

Write $R_i = \mathbb{F}_2[x]/\langle f_i(x)^{e_i} \rangle$.

\begin{Definition}\label{def:8-5}

Assume $A$ is invertible (equivalently $3 \nmid n$, Theorem~\ref{thm:5-2}), so
that $G(x)$ is a unit in $R_n$ and hence in each $R_i$. The \emph{local order}
$k_i$ is the least $k \in \mathbb{N}$ with $G(x)^k \equiv 1
\pmod{f_i(x)^{e_i}}$.

\end{Definition}

\subsection{The LCM Theorem}

\begin{Theorem}\label{thm:8-6}

If $A$ is invertible, $r(A) = \operatorname{lcm}(k_1,\ldots,k_r)$.

\end{Theorem}

\begin{proof}

By Theorem~\ref{thm:7-6}, $r(A)$ is the least $r$ with $G(x)^r \equiv 1
\pmod{x^n - 1}$. By the CRT isomorphism, this holds iff $G(x)^r \equiv 1
\pmod{f_i(x)^{e_i}}$ for every $i$, i.e.\ iff $k_i \mid r$ for every $i$.
Minimality then gives $r(A) = \operatorname{lcm}(k_1,\ldots,k_r)$.

\end{proof}

Consequently, the global order is determined entirely by the local orders on
the CRT components; orbit periods arise from the interaction of the local
cycles via least common multiples.

\subsection{Nilpotent Components in Even Dimensions}

Assume $n$ is even. By Proposition~\ref{thm:8-2}, some factors of $x^n - 1$
occur with multiplicity $e_i > 1$, and the corresponding local rings $R_i =
\mathbb{F}_2[x]/\langle f_i(x)^{e_i} \rangle$ are not fields but local rings
with non-trivial nilpotents. The decomposition results below apply to any local
component in which the evolution generator is invertible, i.e.\ where $G(x)
\not\equiv 0 \pmod{f_i(x)}$; this holds automatically whenever $A$ is
invertible on the whole space, and is assumed throughout this section.

\begin{Definition}\label{def:8-7}

The maximal ideal of $R_i$ is $\mathfrak{m}_i = \langle f_i(x) \rangle$. It
satisfies $\mathfrak{m}_i^{e_i} = 0$, so every element of $\mathfrak{m}_i$ is
nilpotent.

\end{Definition}

\begin{Theorem}[Unipotent--Semisimple Decomposition]\label{thm:8-8}

Every $g \in R_i^\times$ admits a unique decomposition $g = u(1+\nu)$, where
$u$ has odd order and $\nu \in \mathfrak{m}_i$ is nilpotent. This is the
decomposition of a unit into its odd-order and $2$-power-order components.

\end{Theorem}

\begin{proof}

Write $|g| = q \cdot 2^k$ with $q$ odd, so $\gcd(q, 2^k) = 1$. By B\'ezout's
identity there exist integers $a, b$ with

\[
a \cdot 2^k + b \cdot q = 1.
\]

Since $g^{2^k}$ has order exactly $q$ and $g^q$ has order exactly $2^k$, and $a
\cdot 2^k \equiv 1 \pmod q$, $b \cdot q \equiv 1 \pmod{2^k}$ by the B\'ezout
identity above, both $a$ and $b$ are units modulo $q$ and $2^k$ respectively;
hence

\[
u := (g^{2^k})^a \text{ has order exactly } q, \quad w := (g^q)^b \text{ has
order exactly } 2^k.
\]

Since $g = g^{a \cdot 2^k + b \cdot q} = (g^{2^k})^a \cdot (g^q)^b = uw$, and
$R_i^\times$ is abelian, this gives $g = uw$ with $u$ of odd order $q$ and $w$
of order the power of two $2^k$, as required.

Let $d = \deg f_i$, so the residue field $\kappa_i := R_i/\mathfrak{m}_i$
is isomorphic to $\mathbb{F}_{2^d}$, and its multiplicative group
$\kappa_i^\times$ has odd order $2^d - 1$. Reducing $w$ modulo
$\mathfrak{m}_i$, its image lies in $\kappa_i^\times$; since this image also
has order dividing $2^k$, and the only element of $\kappa_i^\times$ whose order
divides both $2^k$ and the odd number $2^d - 1$ is the identity, the image of
$w$ equals $1$. Hence $w = 1 + \nu$ for some $\nu \in \mathfrak{m}_i$, which is
nilpotent since $\mathfrak{m}_i^{e_i} = 0$. Suppose $g = u_1(1+\nu_1) =
u_2(1+\nu_2)$ are two such decompositions, with $u_1, u_2 \in R_i^\times$ of
odd order and $\nu_1, \nu_2 \in \mathfrak{m}_i$. Reducing modulo
$\mathfrak{m}_i$ gives $\bar{u}_1 = \bar{u}_2$ in $\kappa_i^\times$ (since
$1+\nu_j \equiv 1 \pmod{\mathfrak{m}_i}$). Then $z := u_1 u_2^{-1}$ has odd
order (being a quotient of odd-order elements in the abelian group
$R_i^\times$) and $\bar{z} = 1$, so $z \in 1+\mathfrak{m}_i$; by Theorem~\ref{thm:8-9}
below, $z$ therefore has order a power of two, forcing $z = 1$, i.e.\ $u_1 =
u_2$. Since $u_1$ is a unit, $u_1(1+\nu_1) = u_1(1+\nu_2)$ then gives $1+\nu_1
= 1+\nu_2$, so $\nu_1 = \nu_2$.

\end{proof}

The decomposition separates each invertible element into a semisimple part
(odd order) and a unipotent part (power-of-two order).

\subsection{Frobenius Reset}

\begin{Theorem}\label{thm:8-9}

If $\nu \in \mathfrak{m}_i$, then $(1+\nu)^{2^t} = 1$ for sufficiently large
$t$.

\end{Theorem}

\begin{proof}

By the Frobenius identity, $(1+\nu)^{2^t} = 1+\nu^{2^t}$. Since $\nu$ is
nilpotent, $\nu^{2^t} = 0$ for large $t$.

\end{proof}

The unipotent component of any $g \in R_i^\times$ therefore always has order
a power of two; in even dimensions, all contributions to orbit length from
nilpotent structure are dyadic.

\subsection{The Minimal Nilpotent Valuation Lemma}

\begin{Lemma}[Minimal Nilpotent Valuation Lemma]\label{thm:8-10}

Let $n = 2^m q$ with $q$ odd and $m \geq 1$, and let
$R_i = \mathbb{F}_2[x]/\langle f_i(x)^{e_i} \rangle$ be a CRT component with
$f_i(x) \neq x+1$. Then $e_i = 2^m \geq 2$, and if
$G_i = u_i(1+\nu_i)$ is the local semisimple--unipotent decomposition of the
evolution generator (Theorem~\ref{thm:8-8}), where $u_i \in R_i^\times$ is the
semisimple part and $\nu_i \in \mathfrak{m}_i$ is nilpotent, then

\[
\nu_i \in \mathfrak{m}_i \setminus \mathfrak{m}_i^2, \quad
\text{equivalently}\quad v_{f_i}(\nu_i) = 1.
\]

\end{Lemma}

\begin{proof}

Let $\zeta$ denote the image of $x$ in the residue field $\kappa_i :=
R_i/\mathfrak{m}_i$, and let $d = \deg f_i$, so $\kappa_i \cong \mathbb{F}_{2^d}$.
We write $v(\cdot)$ for $v_{f_i}(\cdot)$ throughout this proof.

\paragraph{Construction (the Teichm\"uller lift of $x$).}

Define $y_0 = x$ and $y_{j+1} = y_j^{2^d}$ for $j \ge 0$. We claim this
sequence stabilizes in $R_i$ after finitely many steps.

First, $v(x^{2^d} - x) = 1$: writing $h(y) = y^{2^d-1} - 1$, since $2^d-1$
is odd we get $h'(\zeta) = \zeta^{2^d-2} = \zeta^{-1} \neq 0$, so $\zeta$ is a
simple root of $h$; writing $h(y) = f_i(y) Q(y)$, separability of $f_i$ gives
$Q(\zeta) \neq 0$, so $Q(x)$ is a unit and $v(x^{2^d-1} - 1) = v(f_i(x)) +
v(Q(x)) = 1$. Since $x$ is a unit, $v(x^{2^d} - x) = v(x(x^{2^d-1} - 1)) = 1$.
Write $\delta := x^{2^d} - x$, so $v(\delta) = 1$.

By induction, $y_{j+1} - y_j = \delta^{(2^d)^j}$ for every $j \ge 0$: this
holds for $j=0$ by definition, and if $y_j - y_{j-1} = \delta^{(2^d)^{j-1}} =:
c$, then, using that the map $z \mapsto z^{2^d}$ is additive in characteristic
$2$,

\[
y_{j+1} - y_j = y_j^{2^d} - y_j = (y_{j-1} + c)^{2^d} - (y_{j-1} + c) =
(y_{j-1}^{2^d} - y_{j-1}) + (c^{2^d} - c) = c + (c^{2^d} - c) = c^{2^d}.
\]

Hence $v(y_{j+1} - y_j) = (2^d)^j \cdot v(\delta) = (2^d)^j$, which exceeds
$e_i$ for $j$ large enough, so $y_{j+1} = y_j$ exactly in $R_i$ (since
$\mathfrak{m}_i^{e_i} = 0$). Call this stable value $\tilde{x}$. By
construction $\tilde{x} \equiv x \pmod{\mathfrak{m}_i}$ and $\tilde{x}^{2^d} =
\tilde{x}$.

Moreover, since

$$
v(y_j - y_1) \geq (2^d)^1 = 2^d \geq 2
$$

for every $j \geq 1$ (each successive difference has valuation $(2^d)^j$, $j \geq 1$), we get
\begin{equation}\tag{$\ast$}
\tilde{x} \equiv x^{2^d} \pmod{\mathfrak{m}_i^2}.
\end{equation}

\paragraph{Uniqueness.}

If $\tilde{x}'$ also satisfies $\tilde{x}' \equiv x \pmod{\mathfrak{m}_i}$ and
$(\tilde{x}')^{2^d} = \tilde{x}'$, let

$$
\varepsilon = \tilde{x} - \tilde{x}' \in \mathfrak{m}_i.
$$

By additivity of $z \mapsto z^{2^d}$ in characteristic $2$,

$$
\varepsilon
= \tilde{x} - \tilde{x}'
= \tilde{x}^{2^d} - (\tilde{x}')^{2^d}
= (\tilde{x} - \tilde{x}')^{2^d}
= \varepsilon^{2^d}.
$$

If $\varepsilon \neq 0$ with $v(\varepsilon) = j \geq 1$, then

$$
v(\varepsilon^{2^d}) = 2^d j \neq j
$$

(as $d \geq 1$), contradicting $\varepsilon = \varepsilon^{2^d}$.
Hence $\varepsilon = 0$ and

$$
\tilde{x} = \tilde{x}'.
$$

The same argument, applied verbatim to any $\alpha, \beta \in R_i$ with
$\alpha \equiv \beta \pmod{\mathfrak{m}_i}$ and $\alpha^{2^d} = \alpha, \beta^{2^d} =
\beta$, shows $\alpha = \beta$ --- that is, the fixed point of the $2^d$-power
map in a given residue class modulo $\mathfrak{m}_i$ is unique whenever it
exists. We use this general form of uniqueness, not only the special case
$\alpha = \tilde{x}$, in Step 2 below.

\paragraph{Step 1 ($f_i$ does not divide $G'(x)$).} Since $f_i \neq x+1$,
its order $e = \operatorname{ord}(\zeta)$ is an odd divisor greater than $1$ of
the odd part $q$ of $n = 2^m q$. The evolution polynomial is $G(x) = 1 + x +
x^{n-1}$, whose formal derivative over $\mathbb{F}_2$ is $G'(x) = 1 + x^{n-2}$,
since $n-1$ is odd. Suppose for contradiction that $f_i(x) \mid G'(x)$.
Evaluating at $\zeta$ gives $1 + \zeta^{n-2} = 0$, i.e.\ $\zeta^{n-2} = 1$, so
$e \mid (n-2)$. But $e \mid q$ and $n = 2^m q$ give $n \equiv 0 \pmod e$,
hence $n - 2 \equiv -2 \pmod e$, so $e \mid (n-2)$ forces $e \mid 2$ ---
impossible, since $e > 1$ is odd. Hence $f_i \nmid G'(x)$.

\paragraph{Step 2 ($G(\tilde{x})$ equals the semisimple part $u_i$).}

Since $G$ has coefficients in $\mathbb{F}_2$, the map $z \mapsto z^{2^d}$
commutes with evaluation of $G$, so $G(\tilde{x})^{2^d} = G(\tilde{x}^{2^d}) =
G(\tilde{x})$; thus $G(\tilde{x})$ is itself fixed by the $2^d$-power map.
Since $G_i \equiv G(\zeta) \pmod{\mathfrak{m}_i}$ and, by
Theorem~\ref{thm:8-8}, $G_i \equiv u_i \pmod{\mathfrak{m}_i}$ with $u_i$ of odd
order dividing $2^d - 1$ (so $u_i^{2^d} = u_i$ as well), both $G(\tilde{x})$
and $u_i$ are fixed by the $2^d$-power map and share the same residue
$G(\zeta)$ modulo $\mathfrak{m}_i$. By the uniqueness just proved,
$G(\tilde{x}) = u_i$.

\paragraph{Step 3 (valuation of $x - \tilde{x}$).}

By $(\ast)$, $x - \tilde{x} \equiv x - x^{2^d} = -\delta \pmod{\mathfrak{m}_i^2}$,
and $v(\delta) = 1 < 2$, so $v(x - \tilde{x}) = 1$ exactly.

\paragraph{Step 4 (conclusion).}

The Taylor expansion of $G$ at $\tilde{x}$ gives the exact identity

\[
G(x) - G(\tilde{x}) = G'(\tilde{x}) \cdot (x - \tilde{x}) + (x - \tilde{x})^2
\cdot R(x)
\]

for some polynomial $R$. By Step 1, $G'(\tilde{x})$ is a unit; by Step 3, $v(x -
\tilde{x}) = 1$, so the first term has valuation exactly $1$ while the second
has valuation $\ge 2$. Hence $v(G(x) - G(\tilde{x})) = 1$. Since $G(\tilde{x})
= u_i$ is a unit (Step 2),

\[
\nu_i = \frac{G(x)}{u_i} - 1 = \frac{G(x) - G(\tilde{x})}{G(\tilde{x})},
\]

and dividing by a unit preserves valuation, so $v(\nu_i) = v(G(x) - G(\tilde{x}))
= 1$, i.e., $\nu_i \in \mathfrak{m}_i \setminus \mathfrak{m}_i^2$.

\end{proof}

\subsection{Exact Local Unipotent Orders}

\begin{Theorem}[Odd Case]\label{thm:8-11-odd}

If $n$ is odd, then every CRT component has multiplicity $e_i = 1$, so
$R_i = \mathbb{F}_2[x]/\langle f_i(x)\rangle$ is a field and
$\mathfrak{m}_i = 0$. Consequently $\nu_i = 0$ and
\[
\operatorname{ord}(1+\nu_i)=1
\]
for every $i$.

\end{Theorem}

\begin{proof}

By Proposition~\ref{thm:8-1}, $x^n-1$ is separable, hence every irreducible
factor occurs with multiplicity $e_i=1$. Thus every local ring $R_i$ is a field,
so its maximal ideal is $\mathfrak{m}_i=0$. Therefore the nilpotent component in
the decomposition $G_i=u_i(1+\nu_i)$ is $\nu_i=0$, and
$\operatorname{ord}(1+\nu_i)=1$.

\end{proof}

\begin{Theorem}[Exact Local Unipotent Orders in Even Dimensions]\label{thm:8-11}

Let $n = 2^m q$ with $q$ odd and $m \geq 1$, and let
$R_i = \mathbb{F}_2[x]/\langle f_i(x)^{e_i}\rangle$ be a local CRT component.
The local unipotent order is determined by the irreducible factor $f_i$
together with its multiplicity $e_i$ in $x^n-1$. More precisely,

\[
\operatorname{ord}(1+\nu_i)=
\begin{cases}
1, & f_i=x+1,\ m=1,\\[2mm]
2^{m-1}, & f_i=x+1,\ m\ge 2,\\[2mm]
2^m, & f_i\ne x+1.
\end{cases}
\]

\end{Theorem}

\begin{proof}

We treat the two cases separately.

\textbf{Case 1: the trivial component $f_i = x+1$.} Since $d = \deg(x+1) =
1$, we have $\kappa_i \cong \mathbb{F}_2$ and $\kappa_i^\times$ is trivial.
By Theorem~\ref{thm:8-9}, every element of $1+\mathfrak{m}_i$ has $2$-power
order. Since $u_i$ has odd order and reduces to $1$ in
$\kappa_i^\times=\mathbb{F}_2^\times$, it follows that $u_i=1$. Hence
$G_i=1+\nu_i$, and computing the valuation of $G-1$ directly computes
$v_{x+1}(\nu_i)$.

\textit{Step 1 (the valuation identity).} For $n\ge 3$,
\[
G(x)-1=x(1+x^{n-2}).
\]
We use the standard valuation identity
$v_{x+1}(x^k+1)=2^{v_2(k)}$ for $k\ge1$ (a standard fact about the $2$-adic
factorization of $x^k-1$ over $\mathbb{F}_2$; writing $k=2^a b$ with $b$ odd,
$x^k-1=(x^b-1)^{2^a}$ by Proposition~\ref{thm:8-2}, and $x^b-1$ has a simple
root at $x=1$ since $b$ is odd, giving $v_{x+1}(x^b-1)=1$). Since $x$ is a
unit modulo $x+1$,

\[
v_{x+1}(G-1)=v_{x+1}(1+x^{n-2})=2^{v_2(n-2)}.
\]

\textit{Step 2 (evaluating the $2$-adic valuation of $n-2$).}

If $m=1$ and $q=1$, then $n=2$ and $G(x)=1+x+x^{1}=1$ in $\mathbb{F}_2[x]$.
Hence $\nu_i=0$ and $\operatorname{ord}(1+\nu_i)=1$.

Now assume $m=1$ and $q>1$. Then $n-2=2(q-1)$ with $q$ odd, so $q-1$ is
even and $v_2(n-2)\ge 2$, giving $v_{x+1}(G-1)\ge 4>e_i=2$. Hence
$G\equiv 1\pmod{(x+1)^{e_i}}$, so $\nu_i=0$ in $R_i$ and
$\operatorname{ord}(1+\nu_i)=1$.

If $m\ge 2$: $n-2=2(2^{m-1}q-1)$ with $2^{m-1}q-1$ odd, so
$v_2(n-2)=1$ exactly, and $v_{x+1}(G-1)=2$, independent of $q$. Since
$e_i=2^m\ge4$, the smallest $t$ with $2^t\cdot 2\ge e_i$ is $t=m-1$,
giving $\operatorname{ord}(1+\nu_i)=2^{m-1}$.

\textbf{Case 2: generic components, $f_i\ne x+1$.}

\textit{Step 1 (applying the valuation lemma).} By the Minimal Nilpotent
Valuation Lemma (Lemma~\ref{thm:8-10}), $v_{f_i}(\nu_i)=1$.

\textit{Step 2 (Frobenius squaring).} Since $\mathfrak{m}_i^{e_i}=0$,
successive Frobenius squaring gives
$(1+\nu_i)^{2^t}=1+\nu_i^{2^t}$. Because $v_{f_i}(\nu_i)=1$, we have
$\nu_i^{2^t}=0$ exactly when $2^t\ge e_i$, and
$\nu_i^{2^{t-1}}\neq0$ whenever $2^{t-1}<e_i$. The smallest such $t$ is
$t=\lceil\log_2(e_i)\rceil$, giving
\[
\operatorname{ord}(1+\nu_i)=2^{\lceil\log_2(e_i)\rceil}.
\]

Since $n=2^m q$ with $m\ge1$, every irreducible factor of $x^q-1$ occurs with
multiplicity $e_i=2^m$ in $x^n-1$, so for $f_i\ne x+1$ this becomes
$\operatorname{ord}(1+\nu_i)=2^m$.

\end{proof}

Theorem~\ref{thm:8-11} shows that the trivial component $x+1$ always
contributes strictly less dyadic order than the generic maximum $2^m$ allowed
by its multiplicity — a fact already visible in the $n=4$ case of
Section~\ref{subsec:n4} (Theorem~\ref{thm:6-8}), where the unique local factor is
$x+1$ with $m=2$, and the order of $A$ was found directly to be $2=2^{m-1}$,
not $2^m=4$. This exceptional behaviour of the trivial character, rather than
being a defect, is what makes the global order formula of this section exact.

\subsection{Local Theory: Dyadic Depth and Periods}

\begin{Definition}[Local Nilpotent Depth and Local Dyadic
Depth]\label{def:8-12}

Let $V_n \cong \bigoplus_{i=1}^r R_i$ be the CRT decomposition, $\mathbf{v} =
(v_1, \dots, v_r) \in V_n$, and for each $i$ write $A_i = u_i(1+\nu_i)$ as in
Theorem~\ref{thm:8-8}. The local nilpotent depth of $v_i$ is

\[
\delta_i(\mathbf{v}) = \min\{ k \ge 0 : \nu_i^k v_i = 0 \},
\]

with the convention $\lceil \log_2 0 \rceil := 0$. The local dyadic depth of
$\mathbf{v}$ is

\[
\tau(\mathbf{v}) = \max_{1 \le i \le r} \lceil \log_2 \delta_i(\mathbf{v})
\rceil.
\]

\end{Definition}

\begin{Proposition}[Dyadic Vanishing Criterion]\label{thm:8-13}

For every $i$ and every $t \ge 0$,

\[
A_i^{2^t}(v_i) = u_i^{2^t} v_i \iff 2^t \ge \delta_i(\mathbf{v}).
\]

In particular, the nilpotent contribution to the $i$-th local component
vanishes under $A^{2^t}$ exactly once $t \ge \lceil \log_2 \delta_i(\mathbf{v})
\rceil$.

\end{Proposition}

\begin{proof}

By the Frobenius identity, $(1+\nu_i)^{2^t} = 1+\nu_i^{2^t}$, so

\[
A_i^{2^t}(v_i) = u_i^{2^t}(1+\nu_i^{2^t})v_i = u_i^{2^t}v_i
\iff \nu_i^{2^t}v_i = 0
\iff 2^t \ge \delta_i(\mathbf{v}),
\]

the last equivalence by definition of $\delta_i(\mathbf{v})$.

\end{proof}

\begin{Theorem}[Structure of Orbit Periods]\label{thm:8-14}

Assume $3 \nmid n$, so that $A$ is invertible. Under the notation established
above, with $u_i$ of odd order $m_i$, recall
$\tau(\mathbf{v}) = \max_{1 \le i \le r} \lceil \log_2 \delta_i(\mathbf{v}) \rceil$
(Definition~\ref{def:8-12}). Then:

\begin{enumerate}

\item Each local component has exact period
$L_i(v_i) = 2^{\lceil \log_2 \delta_i(v_i) \rceil}d_i$ for some
$d_i \mid m_i$.

\item The global period is $L(\mathbf{v}) = 2^{\tau(\mathbf{v})} d$, where
$d \mid \operatorname{lcm}(m_1,\ldots,m_r)$.

\item In particular, $v_2(L(\mathbf{v})) = \tau(\mathbf{v})$: the dyadic
valuation of the orbit period is completely determined by the maximal local
nilpotent depth.

\end{enumerate}

\end{Theorem}

\begin{proof}

If $m=0$, then by Proposition~\ref{thm:8-1}, $x^n-1$ is separable, so every
$e_i=1$, every $\mathfrak{m}_i=0$, and every $\nu_i=0$. Hence each local period
is odd (indeed, it is determined entirely by the odd-order element $u_i$), and
for each component
\[
\delta_i(\mathbf{v}) =
\begin{cases}
0, & v_i=0,\\
1, & v_i\ne0.
\end{cases}
\]
Thus $\lceil\log_2\delta_i(\mathbf{v})\rceil=0$ for every $i$, so
$\tau(\mathbf{v})=0$, and statements (1)--(3) follow. We therefore assume
$m\ge1$ for the remainder of the proof.

Since $A$ acts independently on each CRT component,
$L(\mathbf{v}) = \mathrm{lcm}(L_1(\mathbf{v}_1), \dots, L_r(\mathbf{v}_r))$,
so it suffices to determine each local period. Fix a component $i$.

If $v_i=0$, then $L_i(v_i)=1$ and there is nothing to prove. Assume henceforth
that $v_i\ne0$.

If $f_i\ne x+1$, Lemma~\ref{thm:8-10} gives
$\rho_i := v_{f_i}(\nu_i)=1$. If $f_i=x+1$ and $m\ge2$,
Theorem~\ref{thm:8-11} gives $\rho_i=2$. If $f_i=x+1$ and $m=1$,
Theorem~\ref{thm:8-11} gives $\nu_i=0$, so $\delta_i(\mathbf{v})=1$ and
$L_i(v_i)=1$, which is consistent with statement (1). Thus we may assume
$\rho_i=1$ or $\rho_i=2$ from here on.

In either remaining case $R_i = \mathbb{F}_2[x]/\langle
f_i(x)^{e_i}\rangle$ is a finite chain ring with maximal ideal
$\mathfrak{m}_i=(f_i)$. Writing
$\ell := v_{f_i}(v_i)$, one has
\[
v_{f_i}(\nu_i^k v_i)=\ell+k\rho_i
\]
for every $k < \delta_i(\mathbf{v})$, and
$\nu_i^k v_i=0$ once $k\ge\delta_i(\mathbf{v})$.

Write $k=2^s j$ with $j$ odd; by the Frobenius identity,

\[
A_i^{2^s j} = u_i^{2^s j}(1+\nu_i^{2^s})^j.
\]

Suppose, for contradiction, that $A_i^{2^s j} v_i = v_i$ for
some $s$ with $2^s < \delta_i(\mathbf{v})$.

Every nonconstant term in the binomial expansion of
$(1+\nu_i^{2^s})^j v_i$ has valuation at least
$\ell + 2^s\rho_i>\ell$, hence lies in $\mathfrak{m}_i^{\ell+1}$.
Reducing $A_i^{2^s j}v_i=v_i$ modulo $\mathfrak{m}_i^{\ell+1}$ therefore gives

\[
\bar{u}_i^{2^s j} \cdot \bar{v}_i = \bar{v}_i
\quad \text{in } \mathfrak{m}_i^\ell/\mathfrak{m}_i^{\ell+1}
\cong \kappa_i,
\]

where $\bar{u}_i$ is the image of $u_i$ and $\bar{v}_i\neq0$ is the leading
term of $v_i$. Since $\bar{v}_i\neq0$, this forces
$\bar{u}_i^{2^s j}=1$ in $\kappa_i^\times$. Set $w:=u_i^{2^s j}$; then $w$
has odd order, being a power of $u_i$ (Theorem~\ref{thm:8-8}), and
$\bar{w}=1$, so $w\in1+\mathfrak{m}_i$. By Theorem~\ref{thm:8-9}, every element
of $1+\mathfrak{m}_i$ has order a power of two; since $w$ has odd order, this
forces $w=1$, i.e.\ $u_i^{2^s j}=1$ exactly in $R_i$, not merely modulo
$\mathfrak{m}_i$.

With $u_i^{2^s j}=1$, the hypothesis reduces to
$(1+\nu_i^{2^s})^jv_i=v_i$, i.e.

\[
\left[
j\cdot\nu_i^{2^s}
+\binom{j}{2}\nu_i^{2\cdot 2^s}
+\binom{j}{3}\nu_i^{3\cdot 2^s}
+\dots
\right]v_i=0.
\]

Since $j$ is odd, the coefficient of the leading term
$\nu_i^{2^s}v_i$ is odd, hence nonzero in $\mathbb{F}_2$. Moreover,
by the definition of $\delta_i(\mathbf{v})$ and the assumption
$2^s < \delta_i(\mathbf{v})$, we have explicitly
\[
\nu_i^{2^s}v_i\neq 0.
\]
Since $v_{f_i}(\nu_i)=\rho_i$, this term has valuation exactly
$\ell+2^s\rho_i$. Every subsequent term has valuation strictly greater than
$\ell+2^s\rho_i$ (or is already zero), so it lies strictly deeper in the
$\mathfrak{m}_i$-adic filtration.
Since $R_i$ is a chain ring, an element of
valuation exactly $\ell+2^s\rho_i$ cannot be cancelled by a sum of elements of
strictly greater valuation, so the bracketed sum is nonzero --- contradicting
$(1+\nu_i^{2^s})^jv_i=v_i$.

Hence $A_i^{2^s j}v_i\ne v_i$ whenever $2^s<\delta_i(\mathbf{v})$, for every
odd $j$, on every nontrivial local component covered above: no period with
dyadic valuation smaller than
$\lceil\log_2\delta_i(\mathbf{v})\rceil$ can occur.

Let $s_i:=\lceil\log_2\delta_i(\mathbf{v})\rceil$. Then
$\nu_i^{2^{s_i}}v_i=0$, so
$A_i^{2^{s_i}}v_i=u_i^{2^{s_i}}v_i$. Since $u_i^{m_i}=1$, raising both sides to
the $m_i$-th power gives
$A_i^{2^{s_i}\cdot m_i}v_i=v_i$, so
$L_i(v_i)\mid 2^{s_i}m_i$. Combined with the lower bound established above
($v_2(L_i(v_i))\ge s_i$), this gives
$v_2(L_i(v_i))=s_i$ exactly, i.e.\
$L_i(v_i)=2^{s_i}d_i$ for some $d_i\mid m_i$. This gives statement 1.

Since each $d_i \mid m_i$ is odd, the odd part of

$$
L(\mathbf{v}) = \operatorname{lcm}(L_1,\dots,L_r)
$$

divides

$$
\operatorname{lcm}(m_1,\dots,m_r),
$$

while the dyadic part equals the largest dyadic factor among the local periods, namely

$$
2^{\tau(\mathbf{v})}.
$$

This gives statements 2 and 3.

\end{proof}

\subsection{Global Theory: Stabilization and Order}

\begin{Theorem}[Stabilization of Reachable Subspaces]\label{thm:8-15}

Let $W_t = \operatorname{Im}(A^t)$ and
$d(t) = \dim_{\mathbb{F}_2}(W_t)$.
Then:

\begin{enumerate}

\item $V_n = W_0 \supseteq W_1 \supseteq W_2 \supseteq \cdots$, with
$d(0)\ge d(1)\ge \cdots$.

\item There exists $t_0$ with $W_{t_0}=W_{t_0+1}=\cdots$, and
$A|_{W_{t_0}}$ is bijective.

\item Consequently $W_{t_0}=\operatorname{Per}(A)$.

\end{enumerate}

\end{Theorem}

\begin{proof}

Since $A(W_t)=W_{t+1}$, we get $W_{t+1}\subseteq W_t$, hence the descending
chain and the nonincreasing dimension sequence. As $0\le d(t)\le n$ for all
$t$, the sequence stabilizes at some $t_0$; equal dimension together with
inclusion forces $W_{t_0}=W_{t_0+1}$, so $A(W_{t_0})=W_{t_0}$, and
$A|_{W_{t_0}}$ is surjective. Since $W_{t_0}$ is finite-dimensional,
surjective implies bijective.

Every point of $W_{t_0}$ then lies on a finite orbit under an invertible
map, hence is periodic, so $W_{t_0}\subseteq\operatorname{Per}(A)$.
Conversely, if $v\in\operatorname{Per}(A)$, then $v=A^Lv$ for some $L\ge1$,
so $v\in\operatorname{Im}(A^L)=W_L=W_{t_0}$ for $L\ge t_0$. Hence
$\operatorname{Per}(A)=W_{t_0}$.

\end{proof}

\begin{Remark}

This may be interpreted as information dissipation: $W_{t_0}$ is the subspace
of configurations on which no further information is lost, and it coincides
exactly with the periodic part of the system.

\end{Remark}

\begin{Proposition}[Existence of Maximal-Period Orbits]\label{thm:8-16}

Let $A$ be invertible with global order $r(A)$. There exists $v\in V_n$ with
$L(v)=r(A)$.

\end{Proposition}

\begin{proof}

Take $\mathbf{v}=e_0$, i.e.\ $P_0(x)=1$ under $\Phi$. Under the CRT
isomorphism (Theorem~\ref{thm:8-4}), the image of $1$ in each local component
$R_i$ is again the multiplicative identity $1$, so the local trajectory on
component $i$ is $1,G_i,G_i^2,\dots$. By Definition~\ref{def:8-5}, the minimal
$k$ with $G_i^k=1$ is exactly the local order $k_i$, so
$L_i(\mathbf{v}_i)=k_i$ for every $i$. By independence of the CRT components,
\[
L(\mathbf{v})=\mathrm{lcm}(L_1(\mathbf{v}_1),\dots,L_r(\mathbf{v}_r))
=\mathrm{lcm}(k_1,\dots,k_r)=r(A),
\]
the last equality by Theorem~\ref{thm:8-6}.

\end{proof}

We now obtain the main structural theorem of the paper.

\begin{Theorem}[Global Maximal Period Classification]\label{thm:8-17}

Let $n=2^m q$ with $q$ odd and $3\nmid n$.

\textbf{Case $q=1$ (n is a pure power of two).} Here $R_n$ is a single
local ring with trivial factor $f_1=x+1$. Then
\[
r(A)=
\begin{cases}
1, & m\in\{0,1\},\\[2mm]
2^{m-1}, & m\ge2.
\end{cases}
\]

\textbf{Case $q>1$.} Then $x^q-1$ has at least one irreducible factor
besides $x+1$, so the CRT decomposition of $R_n$ contains, alongside the
trivial component $f_1=x+1$, at least one generic component $f_i\ne x+1$.
Let $m_2,\ldots,m_r$ denote the odd orders of the semisimple parts $u_i$ on
these generic components. Then
\[
r(A)=2^m\cdot\operatorname{lcm}(m_2,\ldots,m_r),
\]
and this value is attained by some orbit.

\end{Theorem}

\begin{proof}

By independence of the CRT components,
$r(A)=\operatorname{lcm}(r(A_1),\ldots,r(A_r))$, and by
Theorem~\ref{thm:8-8},
\[
r(A_i)=\operatorname{lcm}\bigl(m_i,\operatorname{ord}(1+\nu_i)\bigr)
\]
for each $i$, since $u_i$ and $1+\nu_i$ commute and have coprime
odd and $2$-power orders.

If $m=0$, then $x^n-1$ is separable by Proposition~\ref{thm:8-1}, so every
local component is a field and every $\nu_i=0$. Hence
\[
r(A)=\operatorname{lcm}(m_1,\ldots,m_r).
\]
For the trivial component $f_1=x+1$, the residue field is $\mathbb{F}_2$, so
$m_1=1$; therefore, when $q>1$,
\[
r(A)=\operatorname{lcm}(m_2,\ldots,m_r),
\]
which is exactly the stated formula because $2^m=1$. When $q=1$, we have
$n=1$, the unique component is $x+1$, and $r(A)=1$.

We now assume $m\ge1$.

If $q=1$, the trivial component is the only component. By
Theorem~\ref{thm:8-11},
\[
r(A)=\operatorname{ord}(1+\nu_1)=
\begin{cases}
1, & m=1,\\[2mm]
2^{m-1}, & m\ge2,
\end{cases}
\]
which gives the stated formula for $q=1$ together with the already handled
case $m=0$.

If $q>1$, at least one generic component $f_i\ne x+1$ exists. Since
\[
x^n-1=(x^q-1)^{2^m}
\]
by Proposition~\ref{thm:8-2}, every irreducible factor of $x^q-1$ occurs
with multiplicity exactly $e_i=2^m$. By Theorem~\ref{thm:8-11}, every generic
component therefore contributes unipotent order
\[
\operatorname{ord}(1+\nu_i)=2^{\lceil\log_2 e_i\rceil}=2^m.
\]
Hence
\[
r(A_i)=\operatorname{lcm}(m_i,2^m)=m_i\cdot2^m
\]
for every generic component, since $m_i$ is odd.

The trivial component contributes
$r(A_1)=\operatorname{ord}(1+\nu_1)$, which equals $1$ when $m=1$ and
$2^{m-1}$ when $m\ge2$ by Theorem~\ref{thm:8-11}. In either case it divides
$2^m$ and so never increases the dyadic factor already supplied by the generic
components. Taking the lcm over all components therefore gives
\[
r(A)=2^m\cdot\operatorname{lcm}(m_2,\ldots,m_r),
\]
proving the formula for $q>1$. That this value is attained follows from
Proposition~\ref{thm:8-16}; combined with Theorem~\ref{thm:8-15}, this is not
merely an upper bound but the exact global order in both cases.

\end{proof}

\begin{Remark}

The case split is necessary: for $q=1$ there is no generic component to
supply the full factor $2^m$, so the trivial component's own (strictly smaller)
order $2^{m-1}$ is the true global order for $m\ge2$ — this is exactly what
Examples~\ref{ex:9-3} and~\ref{ex:9-4} below compute directly. Writing the two
cases as one formula,
$r(A)=2^{m-\varepsilon}\cdot\operatorname{lcm}(m_2,\ldots,m_r)$ with
$\varepsilon=1$ if $q=1$ and $m\ge1$, and $\varepsilon=0$ otherwise
(empty lcm $=1$), is possible but is deliberately avoided here in favour of
explicit cases.

\end{Remark}

\subsection{Summary}

The CRT decomposition reduces the global dynamics to independent local
systems. For odd $n$, all components $R_i$ are fields, contributing cyclic
behaviour of odd order. For even $n$, nilpotent extensions contribute
additional factors, restricted by Theorem~\ref{thm:8-9} to powers of two;
Proposition~\ref{thm:8-13} gives the exact vanishing time of local nilpotent
contributions, Theorem~\ref{thm:8-14} the exact local and global periods, and
Theorem~\ref{thm:8-17} the exact global order.

\section{Finite Field Background and Illustrative Examples}
\label{sec:finite-field}

Section~\ref{sec:crt} reduced the global dynamics to local components via the Chinese Remainder Theorem, and Theorem~\ref{thm:8-17} gives the exact global order: $r(A) = 2^{m-1}$ when $n$ is a pure power of two ($q = 1$), and $r(A) = 2^m \operatorname{lcm}(m_2,\ldots,m_r)$ over the generic components otherwise. This section recalls the finite-field facts underlying the semisimple orders $m_i$, and illustrates the complete theory on several worked examples.

\subsection{Classical Facts}
\label{sec:finite-field-classical}

We use the following standard facts about finite fields.

Recall that for $f(x) \in \mathbb{F}_2[x]$ irreducible of degree $d$, the quotient $\mathbb{F}_2[x]/\langle f(x) \rangle$ is a finite field with $2^d$ elements, whose multiplicative group is cyclic of order $2^d - 1$. By Lagrange's theorem, every nonzero element $g$ satisfies $\operatorname{ord}(g) \mid (2^d - 1)$. If $R_i \cong \mathbb{F}_{2^d}$, it follows that $m_i \mid (2^d - 1)$: all local semisimple orders in separable components divide a number of the form $2^d - 1$.

Recall also that for an odd prime $p$, the degree of every irreducible factor of the cyclotomic polynomial $\Phi_p(x)$ over $\mathbb{F}_2$ equals $\operatorname{ord}_p(2) = \min\{ m \in \mathbb{N} : 2^m \equiv 1 \pmod{p} \}$.

\subsection{Computing Local Orders}

Given the factorization $x^n - 1 = \prod f_i(x)^{e_i}$, computing the local semisimple order $m_i$ on a separable component reduces to computing the multiplicative order of $G(x) \bmod f_i(x)$ inside the finite field $\mathbb{F}_2[x]/\langle f_i(x) \rangle$. In general this requires explicit polynomial arithmetic, as the examples below illustrate; the divisibility $m_i \mid (2^d - 1)$ from Section~\ref{sec:finite-field-classical} narrows the search but does not by itself determine $m_i$.

\subsection{Examples}

\begin{Example}[$n = 5$]\label{ex:9-1}
Here $\operatorname{ord}_5(2) = 4$, so $\Phi_5(x) = x^4+x^3+x^2+x+1$ is irreducible over $\mathbb{F}_2$, and
\[
x^5 - 1 = (x+1)(x^4+x^3+x^2+x+1),
\]
giving $R_5 \cong \mathbb{F}_2 \oplus \mathbb{F}_{16}$. On the trivial component, $G(1) = 1+1+1 = 1$ in $\mathbb{F}_2$, so $k_1 = 1$. On the quartic component, writing $\zeta$ for the image of $x$ (a primitive $5$th root of unity, so $\zeta^5 = 1$), direct reduction using $\zeta^4 = \zeta^3+\zeta^2+\zeta+1$ gives $G(\zeta) = 1+\zeta+\zeta^4 \equiv \zeta^3+\zeta^2$. Computing successive powers modulo $\Phi_5(x)$ (using $\zeta^5 = 1$) gives $G(\zeta)^2 = \zeta^3+\zeta^2+1$ and $G(\zeta)^3 = 1$, while $G(\zeta) \neq 1$; hence $k_2 = 3$. Since $n$ is odd, there are no nilpotent components, and $r(A) = \operatorname{lcm}(1,3) = 3$. This illustrates the purely semisimple case.
\end{Example}

\begin{Example}[$n = 7$]\label{ex:9-2}
Here $\operatorname{ord}_7(2) = 3$, so
\[
x^7 - 1 = (x+1)(x^3+x+1)(x^3+x^2+1),
\]
giving $R_7 \cong \mathbb{F}_2 \oplus \mathbb{F}_8 \oplus \mathbb{F}_8$. As before $k_1 = 1$. On each cubic component, $k_i \mid 7$, and $7$ is prime, so $k_i \in \{1,7\}$. To rule out $k_i = 1$: this would require $G(x) \equiv 1$, i.e.\ $x+x^6 \equiv 0$, i.e.\ $x^5 \equiv 1$ (using $x \neq 0$), modulo $f_i$; since the multiplicative order of $x$ in $\mathbb{F}_8^\times$ is $7$ (prime), $x^5 = 1$ would force $x = 1$, contradicting $f_i \neq x+1$. Hence $k_i \neq 1$, so $k_2 = k_3 = 7$, and $r(A) = 7$: every non-trivial periodic orbit in $V_7$ has length exactly $7$. Like $n = 5$, this illustrates the purely semisimple case, here with both non-trivial components sharing the same order.
\end{Example}

\begin{Example}
\label{ex:6}
[$n = 8$]\label{ex:9-3}
Here $n = 2^3$, so $x^8 - 1 = (x+1)^8$ by Proposition~\ref{thm:8-2}, and $R_8$ is a single local ring with the trivial factor $f = x+1$ at multiplicity $e = 8$. There is no odd part, so $m_i = 1$, and by Theorem~\ref{thm:8-11} (case $f_i = x+1$, $m = 3 \geq 2$), $\operatorname{ord}(1+n_i) = 2^{m-1} = 4$. Hence $r(A) = 4$, strictly less than the trivial bound $n = 8$ from Theorem~\ref{thm:7-10}. This illustrates the exceptional $x+1$ component.
\end{Example}

\begin{Example}
\label{ex:7}
[$n = 16$]\label{ex:9-4}
Here $n = 2^4$, similarly a single trivial component, and by the same case of Theorem~\ref{thm:8-11}, $r(A) = 2^{4-1} = 8$. Like $n = 8$, this illustrates the exceptional $x+1$ component, here at one further multiplicity level.
\end{Example}

\begin{table}[H]
\caption{Summary of the illustrative examples and their exact global orders.\label{tab:examples}}
\centering
\begin{tabularx}{\textwidth}{lCCC}
\toprule
\textbf{$n$} & \textbf{Factorization type} & \textbf{Components} & \textbf{$r(A)$} \\
\midrule
5 & $(x+1)\cdot\Phi_5$ & $\mathbb{F}_2 \oplus \mathbb{F}_{16}$ & 3 \\
7 & $(x+1)\cdot f_1 \cdot f_2$ & $\mathbb{F}_2 \oplus \mathbb{F}_8 \oplus \mathbb{F}_8$ & 7 \\
8 & $(x+1)^8$ & single nilpotent local ring & 4 \\
16 & $(x+1)^{16}$ & single nilpotent local ring & 8 \\
\bottomrule
\end{tabularx}
\end{table}

\paragraph{Comparison with Kaneko~\citep{kaneko2014}.}
The four examples above allow a direct numerical check against Kaneko's results. Example~\ref{ex:9-2} ($n = 7$) is precisely the case $m = 0$, $k = 3$ of his family $N = 2^m(2^k-1)$, i.e., $N = 7 \cdot 2^0 = 7$; his Theorem 3.2 shows this bound is exactly attained, and Theorem~\ref{thm:8-17} independently gives $r(A) = 7$, confirming the match. Examples ~\ref{ex:6}  and ~\ref{ex:7} ($n = 8, 16$) fall in the same family with $k = 1$ ($q = 2^1 - 1 = 1$), for which Kaneko's Theorem 3.1 only guarantees $A^N = I$ without showing $N$ is minimal; Theorem~\ref{thm:8-17} instead gives the exact orders $r(A) = 4$ and $r(A) = 8$ --- strictly smaller than $N = 8$ and $N = 16$ respectively --- confirming that his general bound is not tight outside the $k = 3$ case he treats explicitly. Example~\ref{ex:9-1} ($n = 5$) lies outside Kaneko's family altogether, since $5$ is not of the form $2^k - 1$ for any $k$, illustrating that Theorem~\ref{thm:8-17} applies beyond the cases covered in the prior literature.

\paragraph{Comparison with Tadaki~\citep{tadaki1994}.} Although Tadaki's null (Dirichlet) boundary conditions
make a direct numerical match with the periodic-boundary examples above inappropriate, his
results exhibit a structural dichotomy analogous to the one developed in this paper. Tadaki finds
that the maximum period $\Pi_N$ of Rule 150 under Dirichlet boundary conditions grows
exponentially with $N$ for typical system sizes, but collapses to the much smaller value
$\Pi_N = N+1$ for the exceptional family $N = 2^l - 1$; the steepest peaks he reports, at
$N = 6, 10, 18, 22, 28$, follow $\Pi_N = 2^{N/2+1} - 2$. This mirrors the qualitative shape of
Theorem~\ref{thm:8-17}: generic $n$ admit an order $r(A) = 2^m \cdot \operatorname{lcm}(m_2,\dots,m_r)$
driven by semisimple components and potentially large, whereas the exceptional family $n = 2^m$
collapses to the provably smaller $r(A) = 2^{m-1}$ (Examples ~\ref{ex:6}  and ~\ref{ex:7} ). The specific exceptional
families differ---powers of two under periodic boundary conditions here, versus $N = 2^n - 1$
under Dirichlet conditions in Tadaki's setting---reflecting the differing boundary conditions
rather than a shared underlying mechanism; no quantitative correspondence between the two is
claimed.

The examples illustrate both mechanisms side by side: $n = 5$ and $n = 7$ are purely semisimple, with $r(A)$ an odd divisor of $2^d-1$ for the relevant degree $d$; $n = 8$ and $n = 16$ are purely unipotent, with $r(A)$ a power of two strictly below the trivial bound $n$, consistent with the exceptional behaviour of the $x+1$ component established in Theorem~\ref{thm:8-11}.

\subsection{Consequences}

Orbit lengths are determined by two independent mechanisms: a semisimple component, arising from finite fields, contributing odd factors dividing numbers of the form $2^d-1$ (Section~\ref{sec:finite-field-classical}); and a unipotent component, arising from nilpotent extensions, contributing powers of two (Section~\ref{sec:crt}). The exact global order combining both mechanisms is given by Theorem~\ref{thm:8-17}.

\section{Computational Verification}
\label{sec:computational}

Although every result in Sections~\ref{sec:invertibility}--\ref{sec:finite-field} is established analytically, the finiteness of the configuration space allows the full dynamics to be generated directly for small $n$, giving an independent check of the theory against explicit computation. Four experiments are reported below: the complete state graph for an invertible case ($n = 4$, Section~\ref{sec:comp-n4}) and a non-invertible case ($n = 6$, Section~\ref{sec:comp-n6}), a single maximal-length orbit ($n = 7$, Section~\ref{sec:comp-n7}), and the non-monotone evolution of the Hamming weight along a trajectory (Section~\ref{sec:comp-hamming}). All simulations were implemented in Python using the update rule $A = I + L + R$ over $\mathbb{F}_2$.

\subsection{State Graph for $n=4$}
\label{sec:comp-n4}

The first experiment considers the smallest dimension admitting non-trivial periodic behaviour. Theorems ~\ref{thm:6-4} and ~\ref{thm:6-8} predict that:
\begin{itemize}
    \item Every orbit has period either 1 or 2,
    \item There are no transient states,
    \item The state graph consists only of fixed points and two-cycles.
\end{itemize}

\begin{figure}[htbp]
    \centering
    \includegraphics[width=0.7\textwidth]{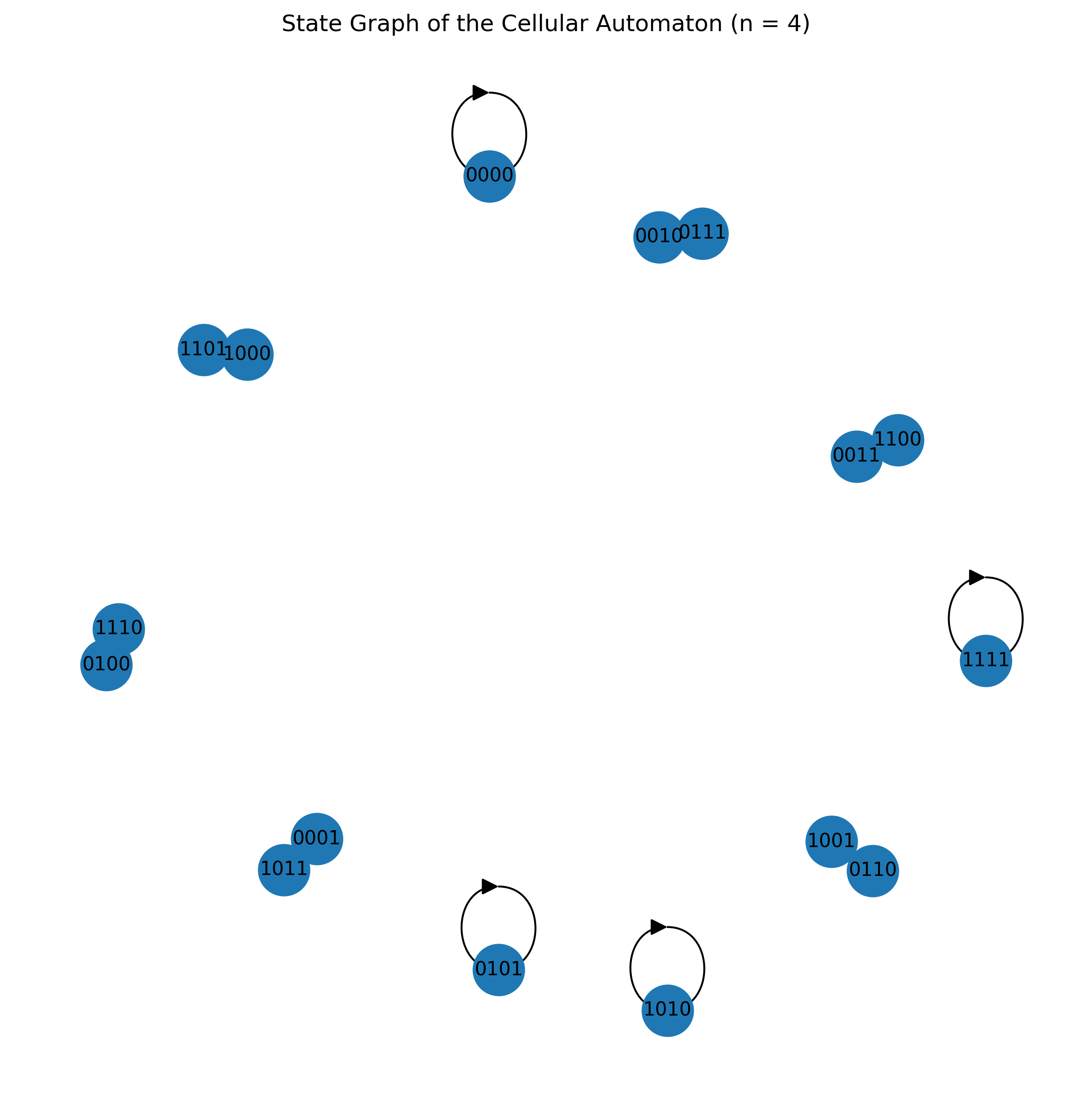}
    \caption{Complete state graph of the automaton for $n = 4$}
    \label{fig:stategraph4}
\end{figure}

To verify this prediction, the complete transition graph of all $2^4=16$ configurations was generated. Figure~\ref{fig:stategraph4} shows the resulting graph.

\paragraph{Discussion.}

The simulation matches the theoretical classification exactly: the graph contains four fixed points and six two-cycles, with no transient trees --- consistent with Proposition~\ref{thm:6-10}, which predicts precisely this count from $|\ker(A \oplus I)| = 4$.

\subsection{State Graph for $n=6$}
\label{sec:comp-n6}

The second experiment illustrates the first non-invertible case in which $3 \mid n$.

\begin{figure}[H]
    \centering
    \includegraphics[width=0.9\textwidth]{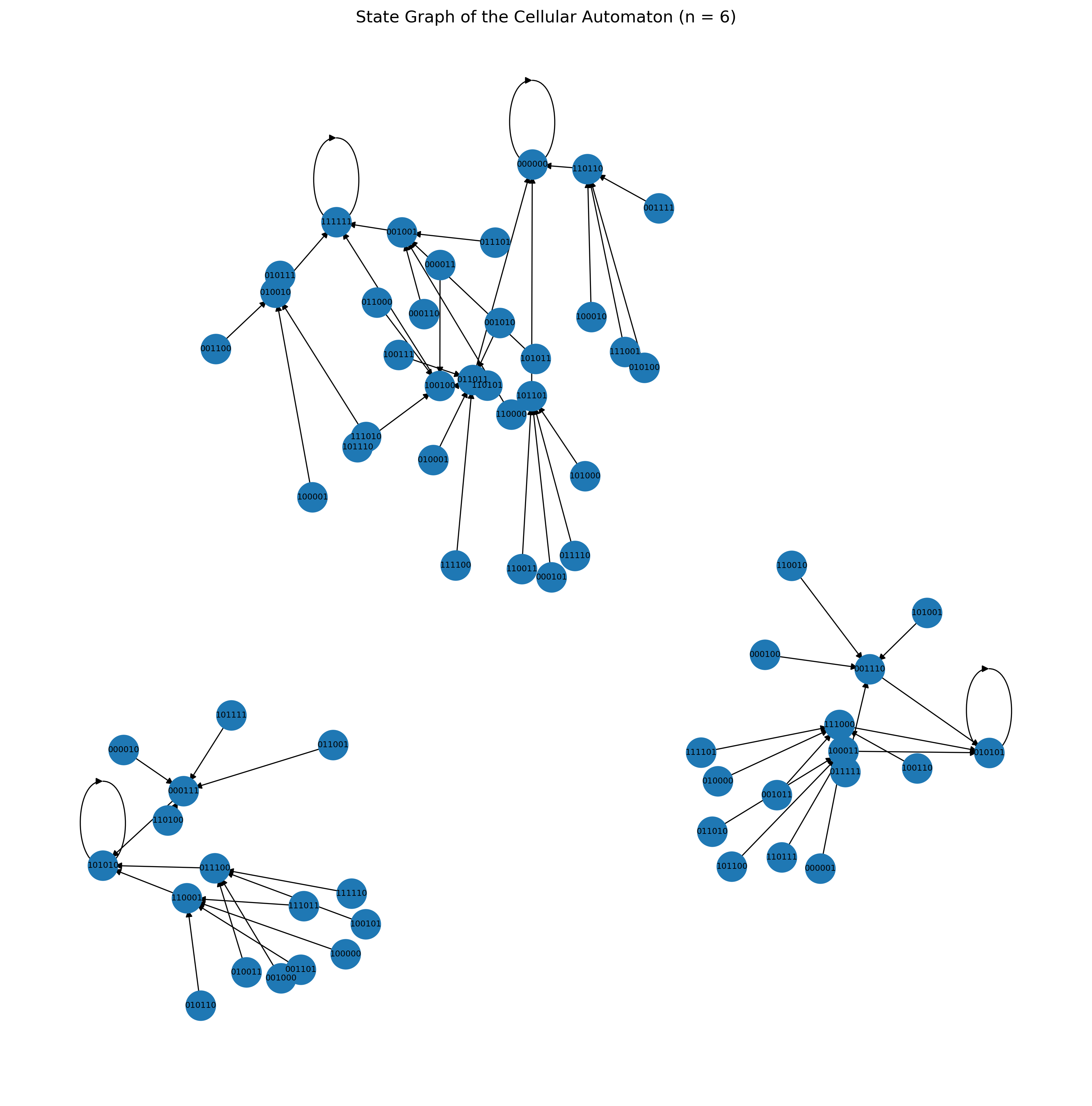}
    \caption{Complete state graph of the automaton for $n=6$, showing transient trees feeding into periodic cycles.}
    \label{fig:stategraph6}
\end{figure}

The complete state graph consisting of all $2^6=64$ configurations was generated. Figure~\ref{fig:stategraph6} displays the resulting dynamics.

\paragraph{Discussion.}

Unlike the case $n=4$, the graph now contains rooted trees feeding into periodic cycles. Multiple configurations merge into common successors, reflecting the loss of invertibility proved in Section~\ref{sec:invertibility}. The transient branches terminate exactly in the periodic core predicted by Theorem~\ref{thm:8-15}. Since $3 \mid 6$, this is the smallest even $n$ for which $A$ fails to be invertible, making it the natural first test case for the transient behaviour classified in Theorem~\ref{thm:5-10}.

\subsection{Example of a Maximal Orbit}
\label{sec:comp-n7}

\begin{figure}[H]
    \centering
    \includegraphics[width=0.95\textwidth]{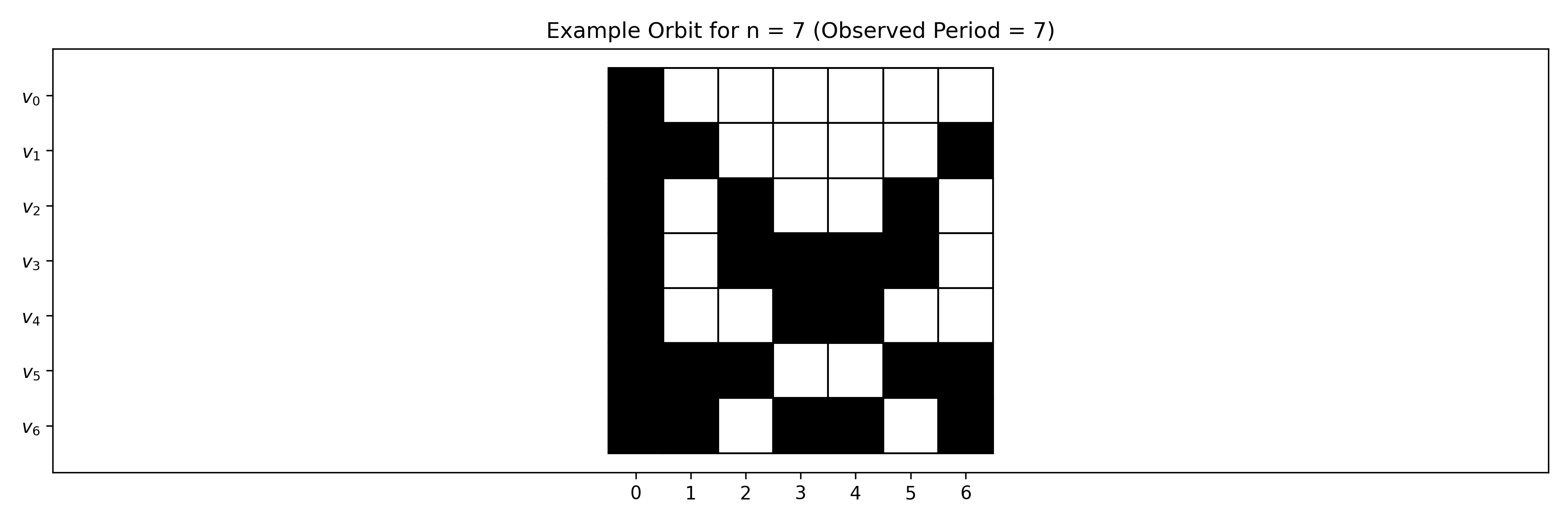}

        \caption{Trajectory of a single initial configuration for $n=7$, iterated until the first repetition.}
    \label{fig:orbit7}
\end{figure}

The third experiment illustrates the behaviour of an individual trajectory in an invertible system. A single initial configuration was selected for $n=7$, and iterated until the first repetition occurred. The observed trajectory is shown in Figure~\ref{fig:orbit7}.

\paragraph{Discussion.}
\begin{figure}

    \label{fig:placeholder}
\end{figure}
The trajectory closes after exactly $7$ steps, matching the value $r(A) = 7$ computed algebraically in Example~\ref{ex:9-2} via the local orders on the two cubic CRT components --- an independent confirmation that the maximal period predicted by Theorem~\ref{thm:8-17} is actually realized by an orbit, as guaranteed by Proposition~\ref{thm:8-16}.

\subsection{Evolution of the Hamming Weight}
\label{sec:comp-hamming}

\begin{figure} [H]
    \centering
    \includegraphics[width=0.9\textwidth]{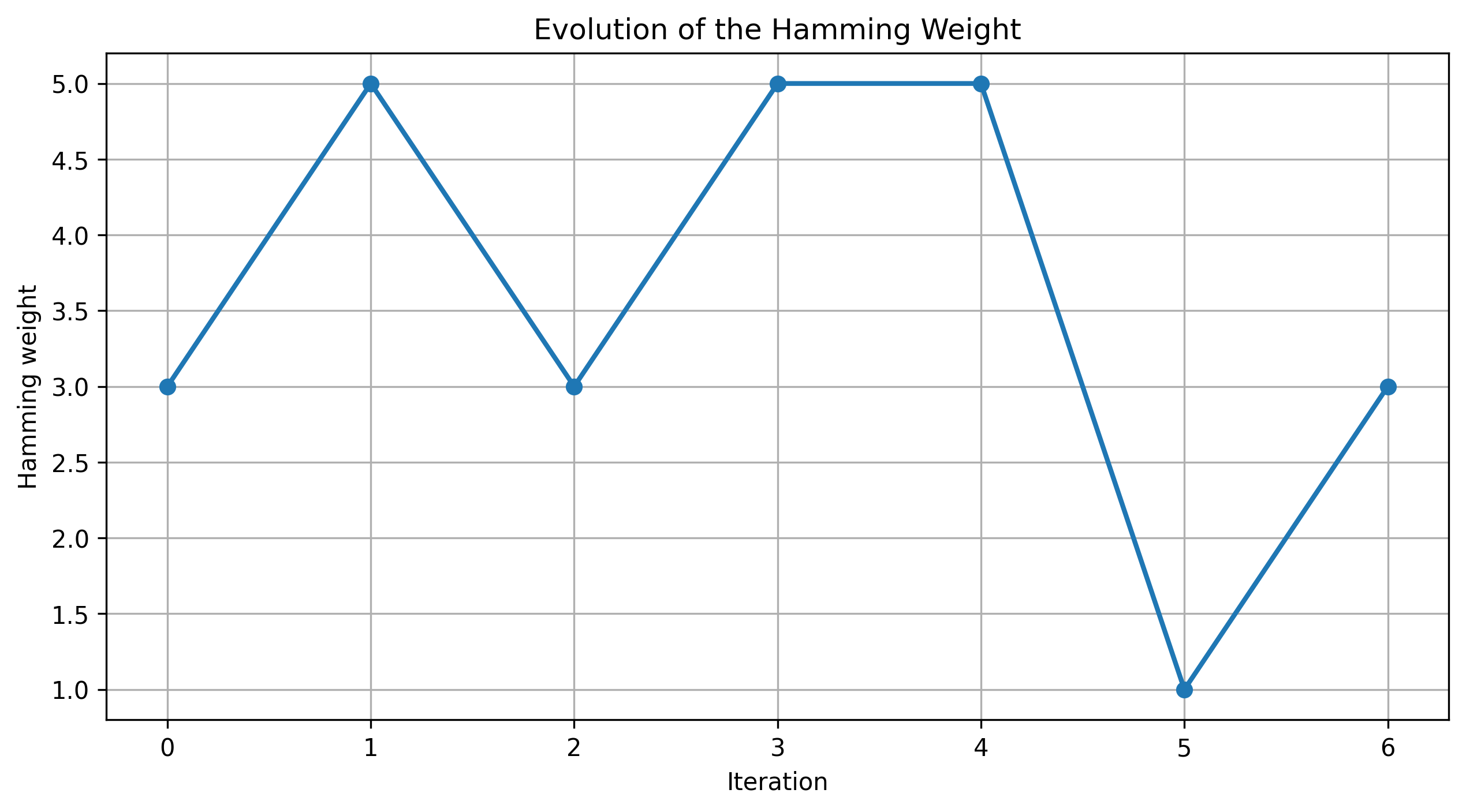}
    
        \caption{Evolution of the Hamming weight along a typical trajectory.}
    \label{fig:hamming}
\end{figure}

The final experiment investigates the Hamming weight
\begin{equation}
w(\mathbf{v}) = \sum_i v_i
\end{equation}
along a typical trajectory. If the Hamming weight were a Lyapunov function, it would necessarily evolve monotonically. Instead, the values shown in Figure~\ref{fig:hamming} fluctuate repeatedly.

\paragraph{Discussion.}

The Hamming weight both increases and decreases during the evolution, exactly as predicted by Theorem~\ref{thm:6-13}: neither $\mathrm{wt}(A\mathbf{v}) \le \mathrm{wt}(\mathbf{v})$ nor $\mathrm{wt}(A\mathbf{v}) \ge \mathrm{wt}(\mathbf{v})$ holds uniformly. Rather than exhibiting dissipative dynamics, the automaton evolves through genuinely cyclic behaviour, confirming that no simple weight-based statistic can serve as a Lyapunov function for this system.

\section{Conclusions and Further Directions}
\label{sec:conclusions}

This paper studied the linear cellular automaton on the cyclic graph $C_n$ over $\mathbb{F}_2$, governed by the local rule $v_i(t+1) = v_{i-1}(t) \oplus v_i(t) \oplus v_{i+1}(t)$. Representing the global evolution operator $A$ as multiplication by $G(x) = 1+x+x^{n-1}$ inside the ring $R_n = \mathbb{F}_2[x] / \langle x^n - 1 \rangle$ (Sections~\ref{sec:algebraic}--\ref{sec:polynomial}) reduced every dynamical question about the automaton---invertibility, orbit structure, periods, transients---to a purely algebraic one.

The central finding of this paper is not any single formula but a structural separation: the dynamics of $A$ decompose, on every CRT component, into a semisimple part governed by multiplicative orders in finite fields and a nilpotent part governed by Frobenius squaring in characteristic $2$, and these two mechanisms interact only through a least common multiple. This separation is what makes the system tractable at all, since it replaces a single opaque question---the long-term behaviour of $A$ on all of $V_n$---with two independent and individually well-understood ones. 

Along the way we obtained a complete invertibility criterion ($A$ fails to be invertible exactly when $3 \mid n$, Section~\ref{sec:invertibility}), a full classification of fixed points, transients, and periodic orbits in terms of this criterion (Sections~\ref{sec:invertibility}--\ref{sec:orbit}), and a proof that the natural candidate Lyapunov function, the Hamming weight, is non-monotone under $A$ for every $n \ge 3$ (Section~\ref{sec:orbit}), ruling out the most obvious energy-like invariant for the system.

The exact global order formula (Theorem~\ref{thm:8-17})---$r(A) = 2^{m-1}$ for $n$ a pure power of two, and $r(A) = 2^m \operatorname{lcm}(m_2,\dots,m_r)$ over the generic CRT components otherwise---together with the exact local unipotent orders of Theorem~\ref{thm:8-11} and the description of how the global order is attained on individual orbits (Proposition~\ref{thm:8-16}), is best understood as a consequence of this separation rather than as its point. It is the semisimple--unipotent decomposition itself---and the arithmetic condition $3 \mid n$, which acts as the single switch between the reversible and irreversible regimes of the system---that gives the theory its explanatory power.

Several directions remain open. First, the exact local unipotent order formula of Theorem~\ref{thm:8-11} was established for the specific generator $G(x) = 1+x+x^{n-1}$ arising from this particular local rule; it would be of interest to determine which of its structural features (the valuation argument of Lemma~\ref{thm:8-10}, in particular) extend to other $\mathbb{F}_2$-linear three-term rules on $C_n$, or to rules on other finite graphs admitting a similar circulant structure. Second, while Section~\ref{sec:orbit} shows that the Hamming weight fails to be monotone, it remains open whether some other natural statistic of a configuration could serve as a genuine (possibly only locally defined, or only strictly monotone off the periodic part) Lyapunov-type invariant for the transient dynamics. Third, the local dyadic depth $\tau(v)$ introduced in Section~\ref{sec:crt} (Definition~\ref{def:8-12}) is defined only in the invertible regime $3 \nmid n$; determining the analogous quantity, and the resulting transient lengths, in the non-invertible regime $3 \mid n$ --- where genuine transients exist (Theorem~\ref{thm:5-10}) --- remains open, and would extend the semisimple--unipotent framework of Section~\ref{sec:crt} across both regimes.

\section*{Acknowledgments}

The author thanks Professor \'Angel Mart\'in del Rey
for comments on an earlier version of the manuscript.


\begin{thebibliography}{999}

\bibitem{bhattacharjee2016}
K.~Bhattacharjee and S.~Das.
Reversibility of $d$-state finite cellular automata.
\textit{Journal of Cellular Automata} 2016, 11(2--3), 213--245.

\bibitem{cattaneo2004}
G.~Cattaneo, A.~Dennunzio and L.~Margara.
Solution of some conjectures about topological properties of linear cellular automata.
\textit{Theoretical Computer Science} 2004, 325(2), 249--271.

\bibitem{cinkir2011}
Z.~Cinkir, H.~Ak{\i}n and \.{I}.~\c{S}iap.
Reversibility of 1D cellular automata with periodic boundary over finite fields $\mathbb{Z}_p$.
\textit{Journal of Statistical Physics} 2011, 143(4), 807--823.

\bibitem{dow1997}
R.~A. Dow.
Additive cellular automata and global injectivity.
\textit{Physica D: Nonlinear Phenomena} 1997, 110(1--2), 67--91.

\bibitem{hernandez2007}
L.~Hern{\'a}ndez Encinas and {\'A}.~Mart{\'\i}n del Rey.
Inverse rules of ECA with rule number 150.
\textit{Applied Mathematics and Computation} 2007, 189(2), 1782--1786.

\bibitem{kaneko2014}
Y.~Kaneko.
A property of the rule 150 elementary cellular automaton.
\textit{arXiv} 2014, arXiv:1401.4779.

\bibitem{lettieri2006}
S.~Lettieri, J.~G. Stevens and D.~M. Thomas.
Characteristic and minimal polynomials of linear cellular automata.
\textit{Rocky Mountain Journal of Mathematics} 2006, 36(3), 1077--1092.

\bibitem{martin1984}
O.~Martin, A.~M. Odlyzko and S.~Wolfram.
Algebraic properties of cellular automata.
\textit{Communications in Mathematical Physics} 1984, 93(2), 219--258.

\bibitem{nobe2004}
A.~Nobe and F.~Yura.
On reversibility of cellular automata with periodic boundary conditions.
\textit{Journal of Physics A: Mathematical and General} 2004, 37(22), 5789--5808.

\bibitem{sato1993}
T.~Sato.
Decidability of some problems of linear cellular automata over finite commutative rings.
\textit{Information Processing Letters} 1993, 46(3), 151--155.

\bibitem{stevens1999}
J.~G. Stevens.
On the construction of state diagrams for cellular automata with additive rules.
\textit{Information Sciences} 1999, 115(1--4), 43--59.

\bibitem{tadaki1994}
S.~Tadaki.
Orbits in one-dimensional finite linear cellular automata.
\textit{Physical Review E} 1994, 49(2), 1168--1173.

\bibitem{vivaldi1994}
F.~Vivaldi.
Cellular automata and finite fields.
\textit{Physica D: Nonlinear Phenomena} 1994, 79(1), 115--131.

\end{thebibliography}
\end{document}